\documentclass[structabstract]{aa}

\usepackage{amsmath,amssymb}
\usepackage{graphicx}
\usepackage{txfonts}
\usepackage{array}
\usepackage{natbib}
\usepackage{hyperref}
\bibpunct{(}{)}{;}{a}{}{,} 

\begin{document}
   \title{Kinematics in galactic tidal tails}
   \subtitle{A source of Hypervelocity stars?}

   \author{Tilmann Piffl\thanks{\email{til@aip.de}}
          \and
          Mary Williams
          \and
          Matthias Steinmetz
          }

   \institute{Leibniz-Institut f\"ur Astrophysik Potsdam (AIP),
              An der Sternwarte 16, 14482 Potsdam, Germany
             }

   \date{Received April 8, 2011; accepted September 26, 2011}

 
  \abstract
   {Recent observations of stars with unusually large radial velocities in the Galactic stellar halo have raised new interest on so-called Hypervelocity stars. Traditionally, it is assumed that the velocities of these stars could only originate from an interaction with the supermassive black hole in the Galactic center. It was suggested that stars stripped-off a disrupted satellite galaxy could reach similar velocities and leave the Galaxy.}
  %
   {In this work we study in detail the kinematics of tidal debris stars to investigate the implications of the new scenario and the probability that the observed sample of Hypervelocity stars could partly originate from such a galaxy collision.}
  %
   {We use a suite of $N$-body simulations following the encounter of a satellite galaxy with its Milky Way-type host galaxy to gather statistics on the properties of stripped-off stars. We study especially the orbital energy distribution of this population.}
  %
   {We quantify the typical pattern in angular and phase space formed by the debris stars. We further develop a simple stripping model predicting the kinematics of stripped-off stars. We show that the distribution of orbital energies in the tidal debris has a typical form which can be described quite accurately by a simple function. Based on this we develop a method to predict the energy distribution which allows us to evaluate the significance and the implications of high velocity stars in satellite tidal debris.}
  %
   {Generally tidal collisions of satellite galaxy produce stars which escape into the intragalactic space even if the satellite itself is on a bound orbit. The main parameters determining the maximum energy kick a tidal debris star can get is the initial mass of the satellite and only to a lower extent its orbit. Main contributors to an unbound stellar population created in this way are massive satellites ($M_{\rm sat} > 10^9$~M$_\odot$). We thus expect intragalactic stars to have a metallicity higher than the surviving satellite population of the Milky Way. However, the probability that the observed HVS population is significantly contaminated by tidal debris stars appears small in the light of our results.}


   \maketitle

\section{Introduction}
The growth of galaxies via the accretion of smaller companion systems is one of the major ingredients in the current perception of galaxy formation and evolution. These satellite galaxies are disrupted in the tidal field of their host galaxies and the new material is dispersed near the orbit of the progenitor. Recent theoretical work has shown that especially the outer stellar halo is predominantly made of stars which were born outside the main galaxy \citep{Abadi2006, Zolotov2009, Scannapieco2011}. Such stars are thought to have low metallicities and to be old. The small fraction of material born inside the main galaxy reaching these large radii was mostly re-distributed during violent major merger events. As these events were more frequent when the galaxy was still young, these stars are also predominantly old. A third small population of the outer halo are the so-called Hypervelocity stars (HVSs) which are ejected via a three-body interaction involving the supermassive black hole (SMBH) in the Galactic center \citep{Hills1988}. Such stars have no age constraints and should be metal-rich as they originate from the innermost region of the galaxy.

The latter population earned attention since they could serve as an indirect proof for the SMBH in the Galactic center \citep{Hills1988} and also because they could be used to measure the shape of the Galactic potential \citep{Gnedin2005, Yu2007, Perets2009}. \citet{Yu2003} estimated  a HVS ejection rate of $10^{-5}$ yr$^{-1}$ and \citet{Perets2007} showed that this rate could increase by a magnitude if massive perturbers such as giant molecular clouds or star clusters were considered.  Aside from the classical ejection mechanism by \citet{Hills1988} several authors have suggested alternative formation scenarios: a binary black hole of equal \citep{Yu2003} and un-equal masses \citep{Levin2006, Sesana2009} or the accretion of a satellite galaxy \citep{Abadi2009}.\\
Recent observations of 17 stars in the Galactic halo with unusually high velocities \citep{Brown2005, Hirsch2005, Edelmann2005, Brown2006a, Brown2006b, Brown2007a, Brown2007b, Brown2009a, Tillich2009} raised new interest on the topic. By design of the search strategy these stars have typically blue colors. They move with velocities up to 720~km~s$^{-1}$ with respect to the Galactic center and are thought to reside at Galactocentric distances of 20-130~kpc. Interestingly, the targeted HVS survey of \citet{Brown2009a} only yield out-going HVSs which is typically attributed to the short lifetimes of the stars compared to the long orbital periods. However, eventually also an in-going star with extremely high velocity was observed \citep{Pryzbilla2010}.\\
Despite the small number of HVSs reported to date several peculiarities in the distribution of the observed population were already claimed. \citet{Abadi2009} found that a large part of the population clusters around a certain travel time ($\sim 130$~Myr), i.e. the time a star would need to travel from the GC and arrive at its current radius with its current radial velocity. Such a clustering is not expected from Hills' original SMBH-ejection scenario. It could be explained, however, by a star burst event near the GC triggering an increased ejection rate of HVSs for certain times.\\
Also the angular distribution on the sky shows signs of anisotropy. \citet{Abadi2009} and \citet{Brown2009b} report a significant overdensity of HVSs around the constellation of Leo. Stars ejected by one or more black holes in the GC should appear on the sky in an approximately homogeneous distribution or in a ring-like structure \citep{Levin2006}. However, a preferred ejection directory as found in the data is not naturally explained with this mechanism (however, see \citet{Lu2010}).

The accreted population of stars in the outer halo can also contain stars with large radial velocities. \citet{Teyssier2009} showed that there should exist an energetically loosely or un- bound population of stars originating from disrupted dwarf galaxies. \citet{Abadi2009} commented that a larger total mass of the Galaxy would allow the normal virialized halo population to reach these velocity regimes. The authors further suggested that the peculiarities in the HVS distribution would be naturally explained if part of the observed HVSs would actually belong to a stream of tidal debris of a recently accreted dwarf galaxy. An example for a HVS likely being generated by this mechanism was recently found in M31 \citep{Caldwell2010}.

Several theoretical studies have already investigated properties of the tidal debris of satellite galaxies. \citet{Johnston1998} approximated the energy distribution of tidal debris particles with a triangular shape to build up a stellar halo distribution. \citet{Choi2007} showed that the energy kick obtained by stripped stars via tidal forces and also the deviations between leading and trailing tidal arms are both increasing with the mass of the approaching satellite. \citet{Warnick2008} investigated the relation between observable properties of tidal streams like radial velocity dispersion or thickness to the properties of the progenitor system. \citet{DOnghia2009, DOnghia2010} investigated the effect of resonances during tidal stripping of rotating systems.

In the present work we investigate the kinematical properties of tidal debris with a special focus on the fastest stars of this population. For this we systematically study tidal encounters of satellite galaxies with their hosts and comprehensively investigate the process. We ran a suite of collisionless $N$-body simulations following the passage of a small companion galaxy through its massive Milky Way-like host galaxy. The set-up of these simulations is described in Sect.~\ref{sec:sim-set-up}. In Sect.~\ref{sec:obs-view} we analyze these simulations to obtain an idea of what properties an observer would find in an HVS population generated in a tidal collision. Then a simple analytical model is developed and tested against the simulations (Sect.~\ref{sec:simple_model}). We show in Sect.~\ref{sec:energyDistribution} how this model can be used to predict the energy distribution of the tidal debris star. In Sect.~\ref{sec:Discussion} we discuss our results and summarize them in Sect.\ref{sec:Summary}.
%
\section{Simulation set-up} \label{sec:sim-set-up}
All simulations were run using the publicly available simulation code Gadget-2 \citep{GadgetPaper}. For the main galaxy we used the model parameters proposed by \citet{Klypin2002}. The galaxy consists of three components, an adiabatically contracted spherical NFW halo \citep{NFW1997}, an exponential stellar disk and a spherical stellar bulge with a Hernquist density profile \citep{Hernquist1990}. The disk density profile is
\begin{equation} \label{eq:rho_disk}
 \rho_{\rm disk}(R,z) = \frac{M_{\rm disk}}{4\pi R_{\rm disk}^2 z_{\rm disk}} \exp\left( -\frac{R}{R_{\rm disk}}\right){\rm sech}^2\left( \frac{z}{z_{\rm disk}}\right),
\end{equation}
where $R=\sqrt{x^2+y^2}$. The vertical scale height is set as $z_{\rm disk}=0.2R_{\rm disk}$, as was found in observations of other disk galaxies \citep{Kregel2002}. Table~\ref{tab:host_parameters} presents the parameters used.

\begin{figure}
  \centering
  \includegraphics[]{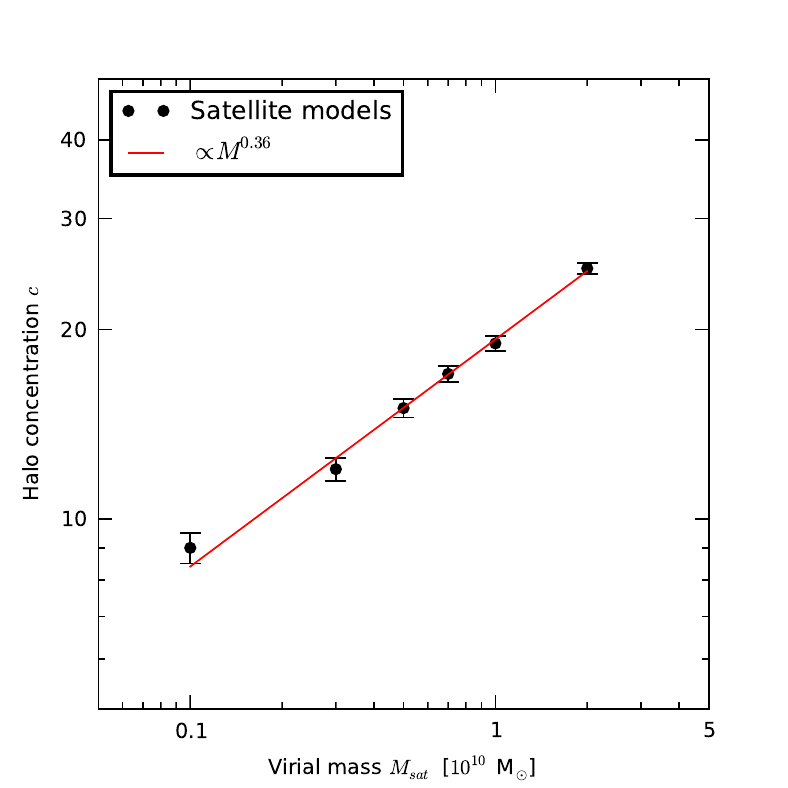}
  \caption{Black symbols show the applied values for the halo concentration parameter $c$ for a given satellite mass $M_{\rm sat}$. The errorbars have a width of 0.5 to reflect the fact that only integer numbers were used for $c$. The red line shows a power law with exponent~0.36. }
  \label{fig:mass-concentration_relation}
\end{figure}
The satellite galaxies are modeled as an adiabatically contracted NFW halo hosting a Hernquist sphere as the spherical baryonic component. The models fulfill the constraints from the Fundamental plane of dE+dSph galaxies given in \citet{deRijcke2005}:
\begin{eqnarray}
  \log L_B &\sim 4.39 + 2.55\log \sigma_0\\
  \log L_B &\sim 8.69 + 3.55\log R_{\rm e}
\end{eqnarray}
Here, $L_B$ is the $B$-band luminosity, $R_{\rm e}$ is the effective radius enclosing half the light of the galaxy and $\sigma_0$ is the luminosity-weighted mean velocity dispersion. A Hernquist sphere has an effective radius $R_{\rm e} \simeq 1.82r_{\rm bulge}$ \citep{Hernquist1990}. For the velocity dispersion $\sigma_0$, in analogy to the dispersion used by \citet{deRijcke2005}, we computed the mass-weighted mean of the line-of-sight velocity dispersions in different radial annuli of the visible component. Finally, to relate the luminosity $L_B$ to the baryonic mass content of the galaxy we assume a mass-to-light ratio $\Upsilon_B = 2\Upsilon_{B,\odot}$.\\
The ratio between total, $M_{\rm sat}$, and baryonic mass, $M_{\rm sat,b}$, was fixed using the relation found by \citet{McGaugh2010}:
\begin{equation}
 \log M_{\rm sat,b} = 4.0\log v_{\rm circ} + 1.65.
\end{equation}
To relate the circular velocity of the dark halo, $v_{\rm circ}$, to the mass of the satellite, $M_{\rm sat}$, we used the same relation as in \citet{McGaugh2010}:
\begin{equation}
 M_{\rm sat} = (1.5\times10^5 {\rm km}^{-3}{\rm s}^3 M_\odot) v_{\rm circ}^3
\end{equation}
With these constraints the dwarf galaxy is fully determined by only one parameter. In this work we used the total mass $M_{\rm sat}$ as a free parameter which was varied for different simulation runs. The requirement to match all the constraints given above fixes also the concentration parameter $c$ of the satellite dark halo. The obtained mass-concentration relation is given in Figure~\ref{fig:mass-concentration_relation}. It is best described by a power law:
\begin{equation}
 c \simeq 19.3 \left(\dfrac{M_{\rm sat}}{10^{10}M_\odot}\right)^{0.36}
\end{equation}
Note, that this relation was obtained by fitting our basic satellite model to the observational constraints. Our concentration parameter should thus not be interpreted in the original sense of an evolutionary sequence in the frame-work of $\Lambda$CDM cosmology \citep{NFW1997}.\\
For the initialization of the phase-space positions of the particle samples we followed a method outlined by \citet{SW1999} which is a modified version of the method of \citet{Hernquist1993} which assumes Gaussian velocity distributions. The latter leads to slight deviations from a perfect equilibrium configuration. To account for this both host and satellite galaxies are allowed to relax for 1~Gyr in isolation before they are implemented into the actual simulations. We use a softening length of 0.01 (0.2) kpc for the satellite star (dark matter) particles in all our simulations.
\paragraph{Simulation time}
All simulations ran until the satellite reached the Apogalacticon after its first pericentric passage or crossed the virial radius of the host galaxy, $R_{200}$. This was done because we wanted to study the properties of a population of tidal debris particles generated during a single stripping event. As we also have a focus on the stars escaping from the combined satellite and host system we chose to study only the first orbit. We expect this orbit to generate the largest spread in velocities as the initial unperturbed satellite population covers the complete possible phase space regions. At later orbits the satellite will have lost its most energetic population \citep{Choi2009}. Furthermore, considering only the first orbit allows an easier comparison between the different simulation runs as one has the full control over the satellite configuration at the beginning of the orbit.
\paragraph{A suite of simulations}
To create a suite of comparable simulations we then ran this scenario with varying initial satellite masses $M_{\rm sat} = 0.1-2 \times 10^{10}$~M$_\odot$ and different starting positions in the satellite phase space. In the majority of cases the satellite is on a bound polar orbit with respect to the host disk component. To determine the influence of an inclined orbit we also ran a couple of simulations with $0^\circ$ (planar) and $45^\circ$ inclination angle. We found that the differences in the results for varying inclinations were small compared to other uncertainties. We thus did not consider inclination as a major parameter and neglect it completely. The initial angular momenta, $L_{\rm sat}$, of the satellites range between 0 and $15 \times 10^3$~kpc~km~s$^{-1}$ which corresponds to pericenter distances $R_{\rm peri}$ from 0 to 50~kpc. Table~\ref{tab:ICs} lists the initial conditions and some of the analysis results for all runs.
\begin{table}
  \centering
  \caption{Parameters of the host galaxy}
  \label{tab:host_parameters}
  \setlength{\extrarowheight}{2pt}
  \begin{tabular}{l l l}
    \hline\hline
    \multicolumn{3}{c}{NFW halo} \\
    \hline
    Total mass, $M_{\rm halo}$ & $113 \times 10^{10}$ & M$_\odot$ \\
    Virial radius, $R_{200}$ & $258$ & kpc \\
    Concentration, $c$ & $12$ & \\
    Virial velocity, $v_{\rm circ}(R_{200})$ & $129$ & km s$^{-1}$ \\
    Particle number, $N_{\rm halo}$ & $5\times10^5$ & \\
    Softening, $\epsilon_{\rm halo}$ & $0.4$ & kpc \\ \hline
    \multicolumn{3}{c}{Exponential disk} \\
    \hline
    Disk mass, $M_{\rm disk}$ & $4.0 \times 10^{10}$ & M$_\odot$ \\
    Scale length, $R_{\rm disk}$ & $3.5$ & kpc \\
    Scale height, $z_{\rm disk}$ & $0.7$ & kpc \\
    Particle number, $N_{\rm disk}$ & $10^5$ &\\
    Softening, $\epsilon_{\rm disk}$ & $0.1$ & kpc \\ \hline
    \multicolumn{3}{c}{Hernquist bulge} \\
    \hline
    Bulge mass, $M_{\rm bulge}$ & $0.8 \times 10^{10}$ & M$_\odot$ \\
    Scale radius, $R_{\rm bulge}$ & $0.7$ & kpc \\
    Particle number, $N_{\rm bulge}$ & $2\times 10^4$ &\\
    Softening, $\epsilon_{\rm bulge}$ & $0.1$ & kpc \\ \hline\hline
  \end{tabular}
\end{table}
%
%

%
\section{Observable properties from the simulations} \label{sec:obs-view}
Of the 41 simulations performed for this study 23 yielded particles with velocities higher than the local escape speed of the host galaxy. These particles are gravitationally unbound and are the simulated equivalent to HVSs. However, in the real Milky Way the escape speed is still uncertain to a considerable degree as neither the total mass \citep[$1-2 \times 10^{12}~{\rm M}_\odot$, e.g.][]{Smith2007, Xue2008, Guo2010, Boylan-Kolchin2011, Przybilla2010} nor the global shape of the gravitational potential \citep{Law2009} is measured accurately. Moreover, the asymmetry of the Galaxy introduces a direction dependency. Thus the value of the escape speed must not be seen as a sharp limiting velocity dividing bound and unbound stars. It is rather a characteristic value to compare to when evaluating the probability whether a star will eventually fall back onto its host or not. For the dynamics of a star located within the virial radius of the Galaxy it makes no qualitative difference whether it is gravitationally unbound. Because of this, and also to obtain better statistics, we will thus in this section analyze the most energetic 0.1\% of the satellite particles, i.e. the Most Energetic Particles (MEPs, for short) regardless whether they reach velocities higher than their local escape speed.
\paragraph{Angular distribution}
\begin{figure*}
  \centering
  \includegraphics{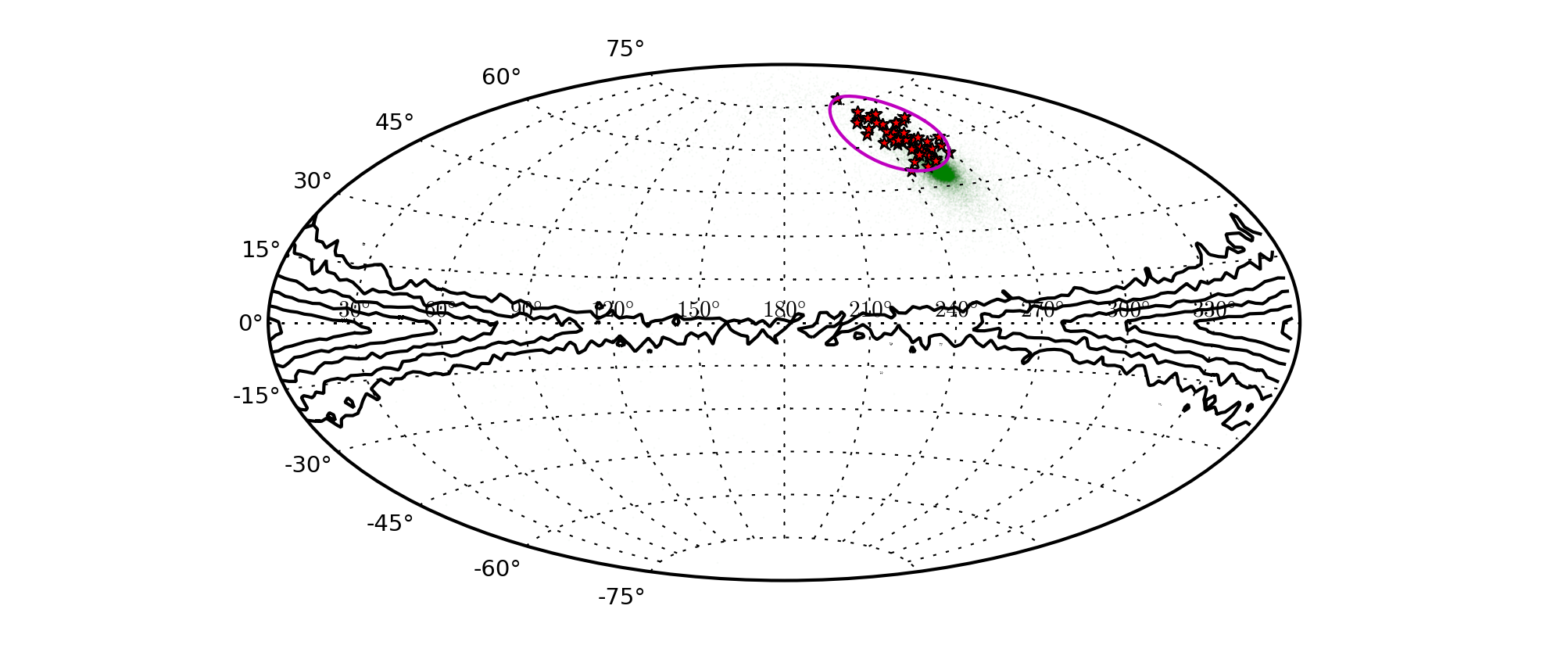}
  \caption{Aitoff projection of a simulation run about 200~Myr after Perigalacticon. Green dots represent all satellite particles (only every 5th particle plotted) while the MEPs are marked by red stars. Disk and bulge of the host galaxy is also shown as black mass-density contour-lines. The median distance of the MEPs is $\sim 60$~kpc, comparable to the Galactocentric distances of the observed HVSs. The MEPs are concentrated to confined area on the sky, in this case enclosed by a circle with angular radius 15$^\circ$.}
  \label{fig:observers-view}
\end{figure*}
Figure~\ref{fig:observers-view} shows the angular distribution of the MEPs in one of our simulations (red stars) as seen by an observer on the Sun's location. The snapshot is taken at a time when the particles have moved from the satellite Perigalacticon out to a galactocentric distance of $\sim 60$~kpc, similar to the observed HVS population. Green dots indicate the positions of all satellite particles while black dots represent star particles belonging to the host galaxy. As already reported by \citet{Abadi2009} the MEPs are clustered in a tightly confined region on the sky (in Fig.~\ref{fig:observers-view} marked by the solid magenta circle of radius $\sim15^\circ$). Averaging over all 39~simulations and over 10 equispaced angular positions of the sun on the solar circle the mean radius of a circular region encompassing all MEP is $16^\circ$ and the maximum angular radius is $27^\circ$. At a distance of 60~kpc the viewing angle of the observer relative to the orbital plane of the satellite does not play a significant role as the stripped-off particles had not enough time yet to unfold into a prominent stream and are thus observable in a compact area from all directions (cf. the upper left panel of Fig.~\ref{fig:RV-dist-plane}).\\
Furthermore, the position of the satellite relative to the MEPs is not arbitrary. In all our simulations the satellite has a smaller angular distance to the host galactic center then the MEPs. This is due to the fact that the satellite as well as the MEPs have just passed their perigalacticon and now move away from the host. Since the MEPs have higher orbital energies they leave the satellite behind and are thus observed at larger angular distances in the vast majority of cases. The angular distance between the MEPs and the satellite remnant is determined by projection effects depending on the viewing angle relative to the orbital plane of the satellite. In our simulations the satellite core is observed always within $22^\circ$ (90~percent within $16^\circ$) separation to the center of the MEP population.
\paragraph{The radial velocity-distance plane}
\begin{figure*}
  \centering
  \includegraphics[scale=1]{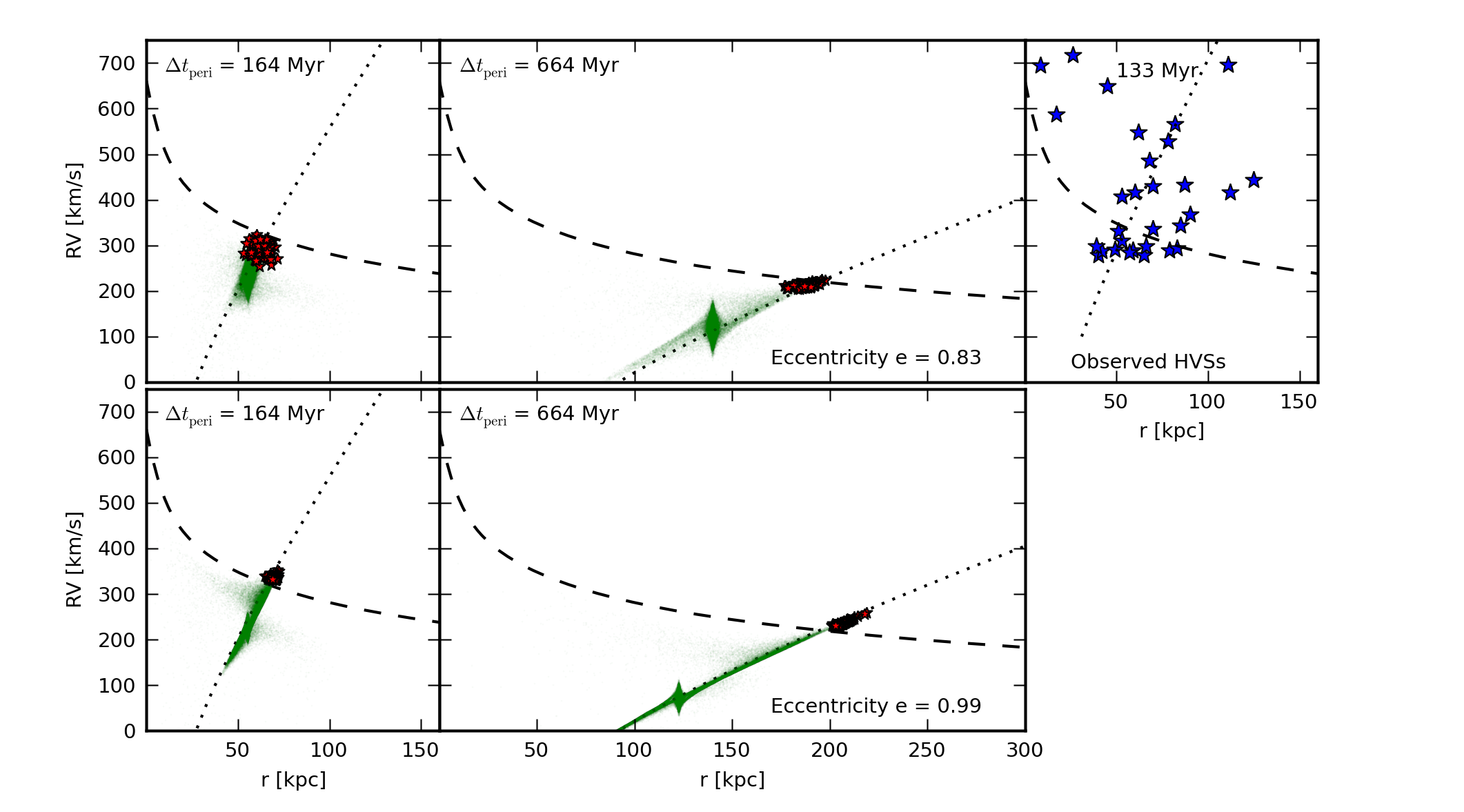}
  \caption{Time series of the galactocentric RV-distance plane for two simulation runs. The time elapsed since Perigalacticon, $\Delta t_{\rm peri}$, is shown in the upper left corner, respectively. Red stars represent the MEPs, green dots satellite particles and dashed lines the local escape speed of the respective radius. Dotted lines mark lines of constant travel times corresponding to $\Delta t_{\rm peri}$. The eccentricity of the satellite orbit is given in the lower right corner of the right most panel, respectively. In the left most panels the mean distance of the MEPs is 60~kpc similar to the observed HVS population which is shown in the right most panel in the upper row for comparison. At this point the MEPs did not have time yet to leave the remaining satellite far behind. Note, that none of the simulations were designed to reproduce the observed HVS population.}
  \label{fig:RV-dist-plane}
\end{figure*}
The particles stripped-off the satellite quickly disperse in physical space and are soon indistinguishable from the already existing Galactic halo population. However, when Galactocentric radial velocities (RVs) are plotted against Galactocentric distance, $r$, the particles form an elongated pattern reflecting their common origin from the Perigalacticon of the satellite. Figure~\ref{fig:RV-dist-plane} shows time series of two simulations with satellite systems on orbits with different eccentricity. Green dots represent satellite particles while red stars show the MEPs. Particles of the host galaxy are not plotted. The upper series is based on the same simulation that was used for Fig.~\ref{fig:observers-view} and the upper left most panel shows the same point of time. The lower panel row shows a simulation run with an almost purely radial orbit which has also lower orbital energy. The initial structural properties of the satellite are the same as in the upper run. Note that despite the lower satellite orbital energy the maximum tidal debris velocities are larger than in the run with a lower eccentricity.\\
The dashed line represents the escape speed from the host system at the respective distance. This corresponds to the trajectory of a test body on a parabolic (purely radial) orbit. Particles below this line are not necessarily bound as part of their motion is hidden in the other velocity components. This explains why particles can cross the dashed line while still conserving their energy.\\
The dotted lines in the panel are lines of constant travel time, i.e. mark positions in the RV-$r$-plane which are occupied by test bodies which started from a common point with varying initial (radial) velocities. The starting point is usually chosen to be the center of the gravitational potential, as in Fig.~\ref{fig:RV-dist-plane}. However, lines from other starting points have similar shapes which explains the good alignment of the tidal debris particles even though the satellite in the upper row run never comes close to the Galactic center (the pericenter distance for this run was 18~kpc).\\
The width of the tidal debris streams in the RV-$r$~plane also changes with eccentricity. This can be understood when considering that most of the stripping happens during a short period of time around the pericenter passage. For more circular orbit this period get more extended as more time is spent by the satellite at radii similar to the pericenter distance.\\
Hence the clustering in travel time already reported by \citet{Abadi2009} has an intrinsic scatter which scales with eccentricity of the orbit of the progenitor system. This is especially so shortly after the pericentric passage, when the distances to the GC are not large and lines of constant travel times have a large slope. At this time the scatter can be stronger than the trend to lines of constant travel time.\\
The left more panel (upper row) shows the distribution of the observed HVSs in the distance-velocity plane for comparison. The possible tidal debris group in the population proposed by \citet{Abadi2009} clusters around the 133 Myr-line of constant travel time show as dotted black line. Note, that none of the two simulation runs was designed to reproduce the observed population as this was not the goal of this more general parameter study.
\paragraph{Maximum velocities}
In the classical SMBH sling shot scenario the extremely large ejection velocities are a result of the extreme orbital velocities occurring near a SMBH plus the large orbital velocities of the components of a hard binary system \citep{Yu2003}. Compared to such an environment, collisions of galaxies are much less violent events as the time scales are much larger and potential gradients much shallower. We thus cannot expect the extraordinary velocities up to 3000 km/s predicted by \citet{Hills1988}. Still, the simulations show that stars can be accelerated to their local escape speed and above. The maximum velocities reached by the MEPs at $r \simeq 60$~kpc in our simulations range between 200 and 400 km/s ($v_{\rm esc}(60{\rm{ }kpc}) = 330$ km/s in our Galaxy model).
%
%
\section{A model for tidal stripping} \label{sec:simple_model}
\begin{figure*}
 \centering
 \includegraphics{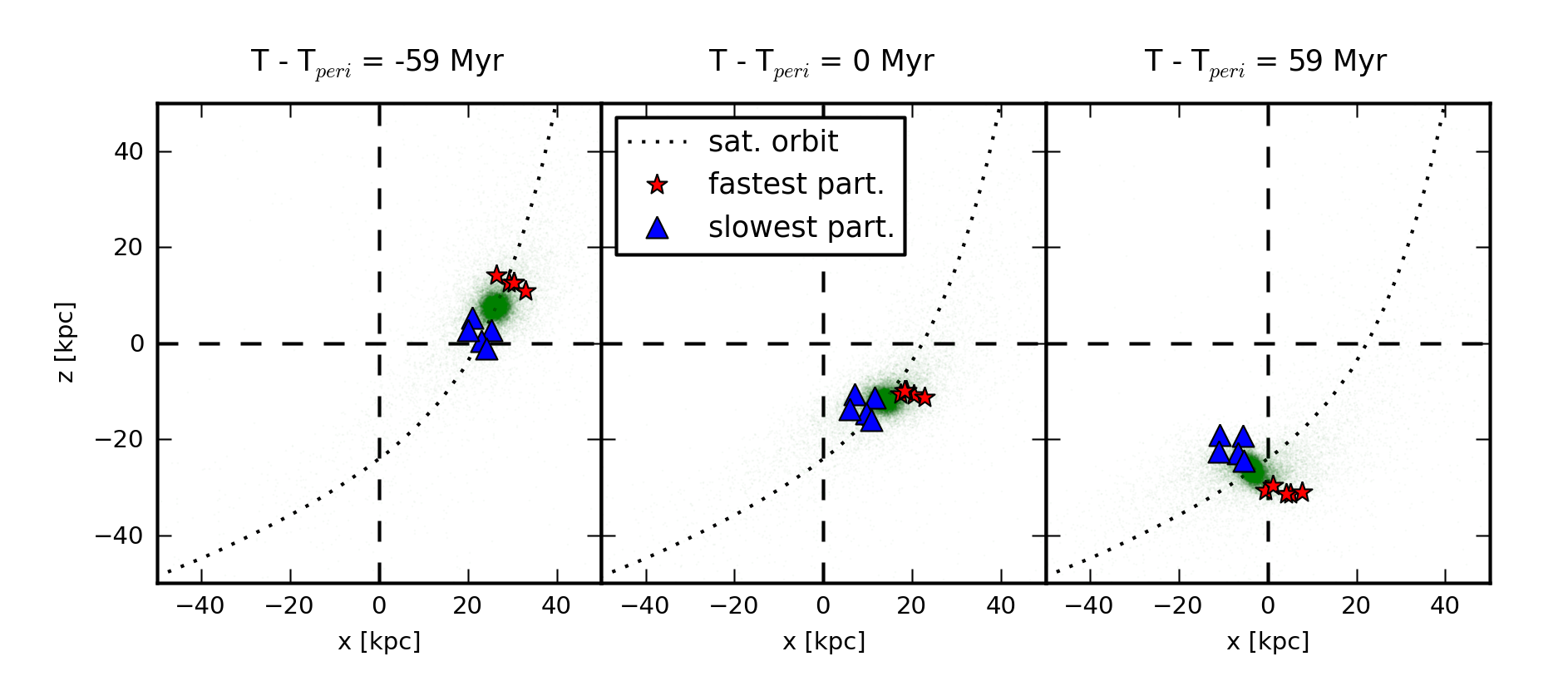}
 \caption{Snapshots of the satellite (small green dots) shortly before, during and shortly after its pericentric passage. Red stars show the positions of those particles which will have the highest orbital energy at the end of the simulation. Blue triangles show those with the lowest energy. The latter have a relative motion opposite to the satellite motion and are on retrograde orbits. The high energy particles move with the satellite and are on prograde orbits.}
 \label{fig:passage_cartoon}
\end{figure*}
\begin{figure}
 \centering
 \includegraphics{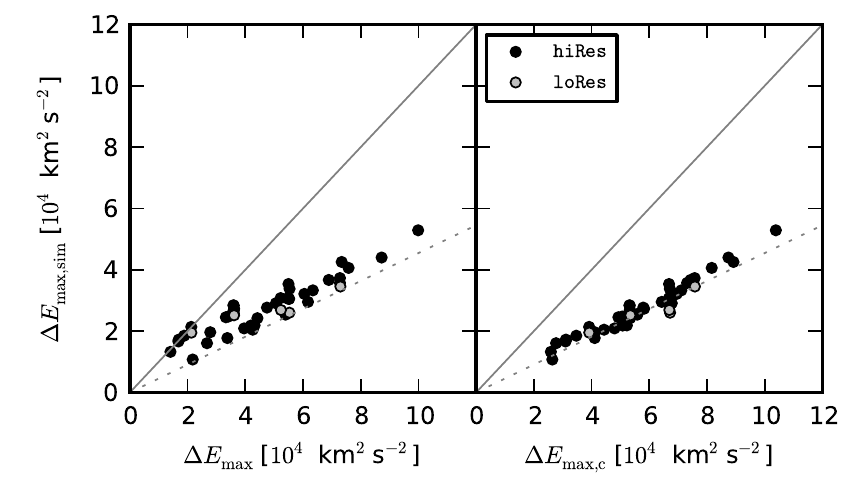}
 \caption{Energy gain with respect to $E_{\rm sat}$ predicted by our simplified model $\Delta E_{\rm max}$ compared to the maximum energy gain found in each simulation. The latter are by a factor of $\sim 0.45$ (slope of the grey dotted lines) lower than the estimates, which is most likely due to oversimplification of the model. Phase-space sampling due to the limited particle number does not play a significant role as can be seen from the grey circles which mark the runs with 10 times lower resolution. {\it Left}: the estimated energy gain as obtained from our model. {\it Right}: the energy gain from our model corrected for an additional dependency on the angular momentum of the satellite. See text for a discussion.}
 \label{fig:dEmax}
\end{figure}
To guide our further analysis we develop a simple, succinct model to describe the accreted satellite mechanism. It was inspired by the calculations of HVS ejection velocities by \citet{Yu2003} as it treats the galaxy-galaxy encounter similar to an binary-SMBH encounter.: a satellite galaxy is moving on an orbit in the gravitational potential $\Phi_{\rm host}(\bf r)$ of its much more massive host galaxy. Its specific orbital energy is thus
\begin{equation}
 E_{\rm sat} = \frac{1}{2}v_{\rm sat}^2 + \Phi_{\rm host}({\bf r}_{\rm sat})
\end{equation}
Since the satellite is an spatially extended object it is subject to tidal forces which lead to a mass loss of the satellite. Under the assumption of an at least moderately eccentric orbit the majority of this stripping will happen in a short period of time when the satellite is closest to the center of the host galaxy where tidal torques are strongest, i.e., at its perigalacticon, $R_{\rm peri}$, where it has the velocity $v_{\rm sat} = V_{\rm peri}$.\\
To model the stripping we now assume what we call \textit{instantaneous escape}: a star $i$ with a position ${\bf r}_{i}$ relative to the satellite center and a velocity ${\bf v}_{i}$ in the co-moving rest frame of satellite has an orbital energy
\begin{equation}
 E_{i} = \frac{1}{2}({\bf v}_{\rm sat} + {\bf v}_{i})^2 + \Phi_{\rm host}({\bf r}_{\rm sat} + {\bf r}_{i}) + \Phi_{\rm sat}({\bf r}_{i}).
\end{equation}
It is lost to the satellite when the gravitational potential from the satellite, $\Phi_{\rm  sat}$ is less than the difference in the host potential between the satellite position and the position of the particle, ${\bf r}_{i}$,
\begin{equation}
 \Phi_{\rm host}({\bf r}_{\rm sat} + {\bf r}_{i}) - \Phi_{\rm host}({\bf r}_{\rm sat}) \ge -\Phi_{\rm sat}({\bf r}_{i}),
\end{equation}
which is equivalent for it to be outside of the tidal radius, $R_{\rm tidal}$. We now assume that this energy transition occurs instantly and the star is left to move in the host potential only. Thus the orbital energy of the star after the stripping is
\begin{eqnarray}
 E_{i}' &=& \frac{1}{2}({\bf v}_{\rm sat} + {\bf v}_{i})^2 + \Phi_{\rm host}({\bf r}_{\rm sat}) = \nonumber\\
        &=& {\bf v}_{\rm sat}{\bf v}_i + \frac{1}{2}v_i^2 + \frac{1}{2}v_{\rm sat}^2 + \Phi_{\rm host}({\bf r}_{\rm sat}) = \\
        &=& {\bf v}_{\rm sat}{\bf v}_i + \frac{1}{2}v_i^2 + E_{\rm sat}.\nonumber
\end{eqnarray}
The energy kick obtained by the star compared to the satellite is then
\begin{equation} \label{eq:deltaEgeneral}
 \Delta E_i = {\bf v}_{\rm sat}{\bf v}_i + \frac{1}{2}v_i^2
\end{equation}
We now ask for the maximum of $\Delta E$. Equation~\ref{eq:deltaEgeneral} leads to the assumption that three conditions need to be fulfilled for the maximum energy kick:
\begin{enumerate}
 \item The star has the maximum velocity possible which is the local escape speed at the tidal radius: $v_i = v_{\rm esc}(R_{\rm tidal})$,
 \item Satellite and stellar velocities have to be aligned: ${\bf v}_{\rm sat} || {\bf v}_i$,
 \item The satellite has to be at its maximum velocity, which occurs during the passage of the Perigalacticon: $v_{\rm sat} = V_{\rm peri}$.
\end{enumerate}
Moreover, if the star is to gain orbital energy it needs to be at larger Galactocentric radii than the satellite, i.e. $|{\bf r}_{\rm sat} + {\bf r}_i| > r_{\rm sat}$, because only then the tidal force push the star away from the galactic center and thus from the potential well. This, together with the first condition, requires the star to be on a prograde orbit with respect to the satellite motion. This view is also confirmed by Fig.~\ref{fig:passage_cartoon}. It shows three snapshots of a simulation run shortly before, at and shortly after Perigalacticon. Red stars and blue triangles mark those particles which will have the highest/lowest orbital energy at the end of the simulation, respectively. The two groups are situated are very distinct locations with respect to the satellite. The particles which gain energy move along with the satellite on an orbit prograde with respect to the satellite motion while for particles which lose energy the orbital phase is such that their motion is contrary to the direction of the system velocity.\\
Thus we arrive at
\begin{equation} \label{eq:deltaE}
 \Delta E_{\rm max} = V_{\rm peri}v_{\rm esc}(R_{\rm tidal}) + \frac{1}{2}v_{\rm esc}(R_{\rm tidal})^2.
\end{equation}
In the left panel of Fig.~\ref{fig:dEmax} we compare this model prediction obtained from Equation~\ref{eq:deltaE} to simulation results. For the latter we analyzed our simulations at Apogalacticon after the satellite's passage through the host system and the particle with the highest orbital energy ${\rm max}(E_i)$  was identified. The energy gain $\Delta E_{\rm max,sim}$ is then defined as ${\rm max}(E_i) - E_{\rm sat}$. For further details how this estimate was done we refer the reader to the Appendix~\ref{app:EnergyEstimate}. Solid lines in Fig.~\ref{fig:dEmax} indicates equality. Equation~\ref{eq:deltaE} \textit{systematically} overestimates the maximum energy kick by a (constant) factor of $2.2$, as indicated by the dashed line with slope $\alpha = 2.2^{-1} \simeq 0.45$. The overestimation are most likely due to oversimplification of our estimate, e.g. the neglection of the structural evolution of the satellite in the tidal field. Surprisingly, the mass resolution only plays a minor role in the sense that simulation runs with lower particle numbers do not yield significantly smaller maximum energy differences. This is most likely due to the steep slope of the energy distribution of the tidal debris stars (see Sect.~\ref{sec:energyDistribution} and Fig.~\ref{fig:ScTdist}).\\
Much of the scatter in the left panel of Fig.~\ref{fig:dEmax} turned out to be a residual dependency on the initial angular momentum of the respective satellites in the simulation. We obtain a tighter relation (right panel) when we use
\begin{equation} \label{eq:correcteddEmax}
  \Delta E_{\rm max,c} = \Delta E_{\rm max} \sqrt{\frac{L_{\rm sat} + L_{\rm char}}{L_{\rm char}}},
\end{equation}
with $L_{\rm char} = 6300$~kpc~km~s$^{-1}$. Thus the value $\alpha\Delta E_{\rm max,c}$ gives a robust estimate of the maximum energy gain occurring during a satellite-host galaxy encounter.
\paragraph{Dynamical friction} \label{para:dynFriction}
In the course of its orbit the satellite galaxy is also subject to dynamical friction. This means that it will sink deeper into the potential well of the host system. By the time it reaches its Perigalacticon stripped-off stars might have to get over an energy gap much wider to become HVSs. The extent of the energy loss depends mainly on the mass $M_{\rm sat}$ and the orbit of the satellite, in a way which is counterproductive to the ejection energies $\Delta E_{\rm max}$: a more massive satellite ejects stars at higher energies as its internal velocities are larger. This energy gain for the stripped-off stars goes roughly spoken with $\sqrt{M_{\rm sat}}$ since $v_{\rm esc}(r) \propto \sqrt{M_{\rm sat}}$ which goes into our estimate (Eq.~\ref{eq:deltaE}). On the other hand a higher mass results in stronger dynamical friction which roughly goes $\propto M_{\rm sat}$ \citep{Chandrasekhar1943}. In fact in our simulations we find that the loss in orbital energy after one orbit is
\begin{equation} \label{eq:dE_DF}
  \Delta E_{\rm DF} = 2 \times 10^{-4} \left(\frac{M_{\rm sat}}{\rm M_\odot}\right)^{0.78} \left(1+\frac{L_0}{5900 {\rm ~kpc~km s}^{-1}}\right)^{-1},
\end{equation}
which reflects the dependency of the physical extent of the satellite on its mass and also the change of the orbital trajectories with changing orbital angular momentum. We use this simplistic approach as it covers best the effect of a possible reaction of the host galaxy on the intruder (however, see \citet{Taylor2001} and \citet{Gan2010} for a more elaborated approach to model dynamical friction).\\
Thus above a certain mass the satellite will not be able to eject any HVSs since the energy loss of the whole system is larger than the energy gain of the single stars. However, judging from our simulations this will only happen at masses $> 10^{11}$~M$_\odot$. Extrapolating our results into this (major merger) mass regime is not meaningful as a massive intruder will significantly perturb the host galaxy. We can thus just state that this scenario does not occur for minor mergers in the present work.
\paragraph{Hypervelocity stars}
In the context of hypervelocity stars which are unbound to the total system we have to compare $\Delta E$ with $E_{\rm sat}$, the orbital energy of the satellite, since the energy $E$ of such an unbound object must be
\begin{equation}
 E = E_{\rm sat}- \Delta E_{\rm DF} + \Delta E \ge 0.
\end{equation}
Thus the condition
\begin{equation}
 \alpha\Delta E_{\rm max,c} \ge -E_{\rm sat}+\Delta E_{\rm DF}.
\end{equation}
must be fullfilled to allow unbound stars to be generated during such a satellite-host galaxy encounter with $\alpha \simeq 0.45$.
%
%
%
\section{The energy distribution} \label{sec:energyDistribution}
\begin{figure}
  \centering
  \includegraphics{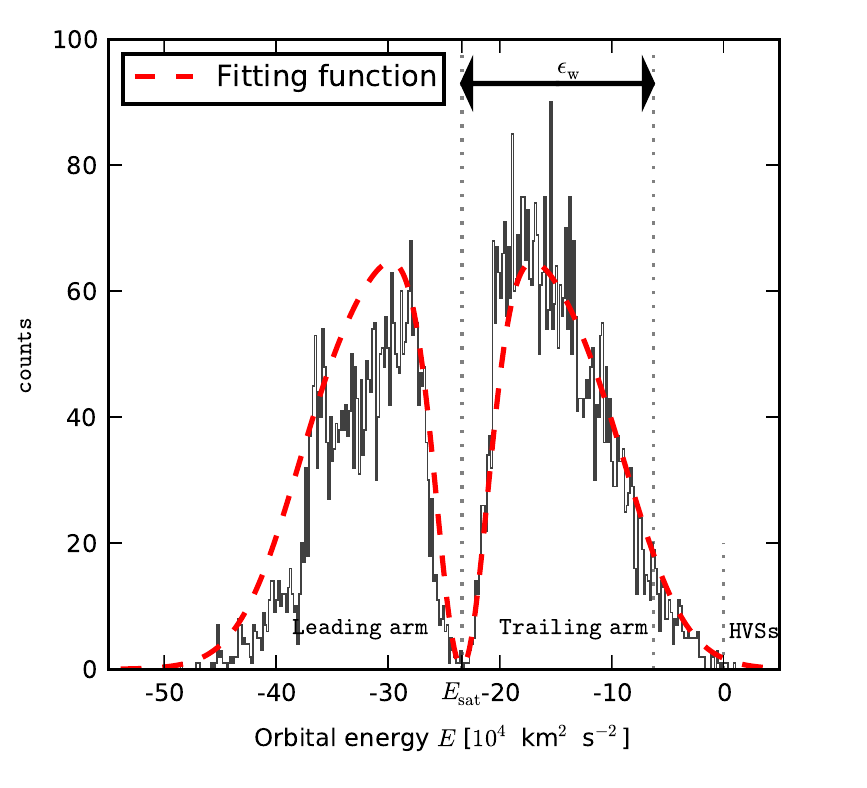}
  \caption{A histogram of the orbital energies of all particles which initially belonged to the satellite but became unbound to it in the course of the first orbit (tidal debris particles). The two peaks correspond to the two tidal arms torn out of the satellite. The central gap coincides with the orbital energy $E_{\rm sat}$ of the remaining satellite system. The dashed rad line shows the fitting function (Eq.~\ref{eq:fitting_function}) which is used to characterize the distribution. The meaning of the fitting parameter $\epsilon_{\rm w}$ which is used to determine the width of the high energy peak is also indicated. Note, that throughout this work we only consider the high energy peak, as this where HVSs would reside.}
  \label{fig:Edist}
\end{figure}
A more robust measure of range of orbital energies covered by the stripped-off stars can be obtained when looking on the overall width of the energy distribution of all particles gravitationally unbound to the satellite. We determine these stars using an iterative method described by \citet{Tormen1998}. The energy distribution is plotted in Fig.~\ref{fig:Edist}. The two peaks represent the leading and trailing tidal arm, respectively. The gap between them is centered on the orbital energy of the satellite. We apply an empirical fitting function
\begin{equation}  \label{eq:fitting_function}
  f_{\rm fit}(E) = \dfrac{f_{\rm high}}{C} \left[\dfrac{1}{(1+\exp(\gamma(\frac{\Delta E}{\epsilon_{\rm w}}-1)))^2} - f_{\rm inner}\exp \left( - \left( \frac{\Delta E}{\epsilon_{\rm inner}} \right) ^2 \right) \right],
\end{equation}
again with $\Delta E = E - E_{\rm sat}$.  To normalize the function we use the factor $f_{\rm high}/C$, where $f_{\rm high}$ is the number of particles in the high energy (trailing) arm divided by the initial number of satellite particles and
\begin{equation} \label{eq:normalization_factor_C}
  \begin{array}{ll}
    C &= \frac{1}{f_{\rm high}}\int_{E_{\rm sat}}^\infty  f_{\rm fit}(E){\rm d}E \\
      &= \epsilon_{\rm w} \left[ 1 + \dfrac{1}{\gamma}\left(\ln(1 + {\rm e}^{-\gamma}) - \frac{1}{1+\exp(-\gamma)}\right)\right] -  \dfrac{\sqrt{\pi}f_{\rm inner}\epsilon_{\rm inner}}{2}.
  \end{array}
\end{equation}
The function is shown in Fig.~\ref{fig:Edist} as red dashed line. For the fitting procedure we only consider the high energy peak of the distribution, i.e. where $E \ge E_{\rm sat}$. The fitting parameters provide us with some characteristics of respective distribution: the width or typical energy, $\epsilon_{\rm w}$, the width of the central minimum, $\epsilon_{\rm inner}$, a measure of how fast the distribution drops off with increasing energy, $\gamma$, and the relative depth of the central minimum, $f_{\rm inner}$.
\paragraph{A composite distribution}
\begin{figure}
  \centering
  \includegraphics{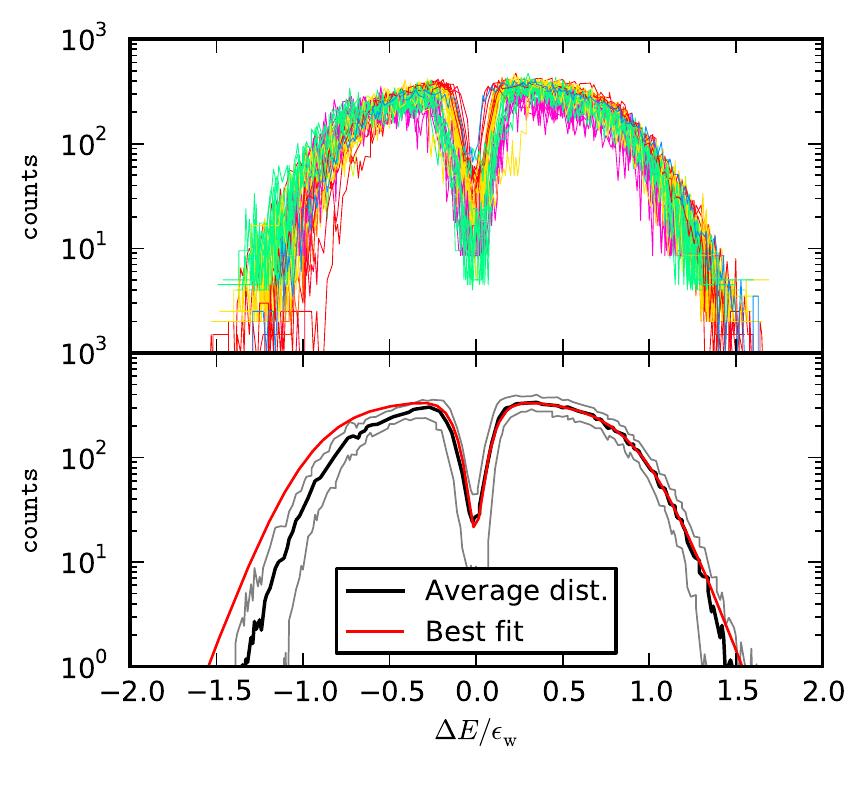}
  \caption{\textit{Upper panel}: The histograms of the energies differences $\Delta E = E_i - E_{\rm sat}$ of the tidal debris particles of all simulations used in this work overplotted. The distributions are plotted as functions of $\Delta E$ in units of the width of the high energy peak $\epsilon_{\rm w}$. The histograms are also renormalized so that the high energy peak covers the same area. The lines are color-coded according to the initial angular momentum of the progenitor system (from red representing more radial orbits to blue for more circular orbits)  \textit{Lower panel}: The mean distribution obtained from the distribution plotted in the upper panel (solid black line). The standard deviation is also indicated with the thin grey lines. Overplotted in red is our fitting function using the parameters given in Eqs.~\ref{eq:FitParams} and $\epsilon_{\rm w} = 1$.}
  \label{fig:AverageDistribution}
\end{figure}
\begin{figure}
  \centering
  \includegraphics{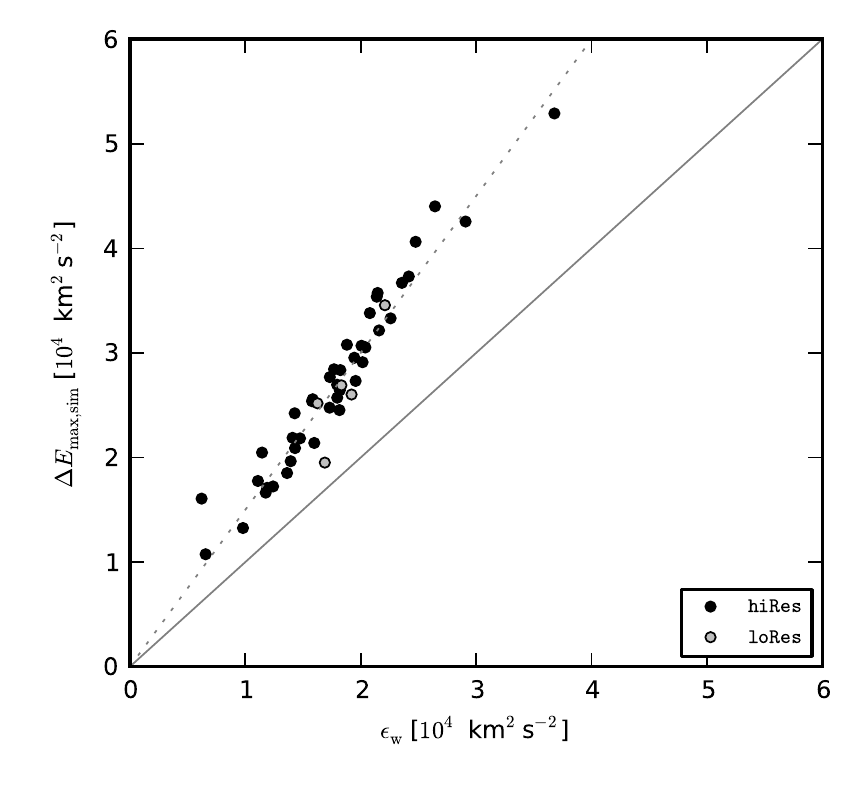}
  \caption{Relation between the maximum energy gain $\Delta E_{\rm max,sim}$ and the width $\epsilon_{\rm w}$ of the energy distribution of all tidal debris stars obtain via fitting Equation~\ref{eq:fitting_function}. The grey dotted line has a slope of 1.5.}
  \label{fig:Emax-Etyp}
\end{figure}
Following an idea by \citet{Johnston1998} we assume the general shape of the distributions to be invariant and only the width and the normalization to be specific to the respective orbit and satellite parameters. This means that the parameters $\epsilon_{\rm inner},\gamma,f_{\rm inner}$ are either constant for all situations or a function of $\epsilon_{\rm w}$. To compare the distribution shapes we rescaled the particle energies into units of their specific typical energy $\epsilon_{\rm w}$ via
\begin{equation}
  \Delta\Hat{E}_i = \frac{E_i - E_{\rm sat}}{\epsilon_{\rm w}}.
\end{equation}
The resulting energy distributions were then renormalized to eliminate the influence of the number particles in the respective (trailing) tidal arm. The upper panel of Fig.~\ref{fig:AverageDistribution} overplots the energy distributions of all our simulations. Note, that these include satellite varying over more than a magnitude in mass and angular momentum. The resulting mean distribution (black line) with the standard deviation (grey line) is also plotted. Applying our fitting function Eq.~\ref{eq:fitting_function} results in the following parameter values:
\begin{eqnarray} \label{eq:FitParams}
  \epsilon_{\rm inner} &=& 0.14 \epsilon_{\rm w}, \nonumber \\
  \gamma &=& 5.21,\\
  f_{\rm inner} &=& 0.94 .\nonumber
\end{eqnarray}
The fit is shown as a dashed red line in Fig.~\ref{fig:AverageDistribution}. We then repeat the fitting procedure for all single distributions with only $\epsilon_{\rm w}$ as a free parameter. The resulting widths are plotted in Fig.~\ref{fig:Emax-Etyp} against the highest energy of all satellite particles, $\Delta E_{\rm max,sim}$. Grey dots represent the lower resolution runs. The tight correlation
\begin{equation} \label{eq:dEmax-A0}
  \Delta E_{\rm max, sim} = \beta \epsilon_{\rm w}
\end{equation}
with $\beta \simeq 1.5$ (which is not strongly affected by the resolution in the simulations) is a result of the steep drop at the high energy tip of the distribution. Using the simple model developed in the previous section we can thus obtain an estimate for the width of the high energy peak via
\begin{eqnarray}
 \epsilon_{\rm w} &=& \beta^{-1} \Delta E_{\rm max,sim} \simeq \frac{\alpha}{\beta} \Delta E_{\rm max,c} \nonumber\\
    &=& 0.3 \Delta E_{\rm max,c}.
\end{eqnarray}
Note, that $v_{\rm esc}(r)$ is a proxy for the mass of the satellite. Thus more massive satellites will produce a larger energy spread in the stripped-off stars. One could also say that the higher velocity dispersion of a more massive galaxy directly translates into a larger energy dispersion in the tidal debris.
%
%
%
%
\section{Discussion} \label{sec:Discussion}
\begin{figure}
  \centering
  \includegraphics{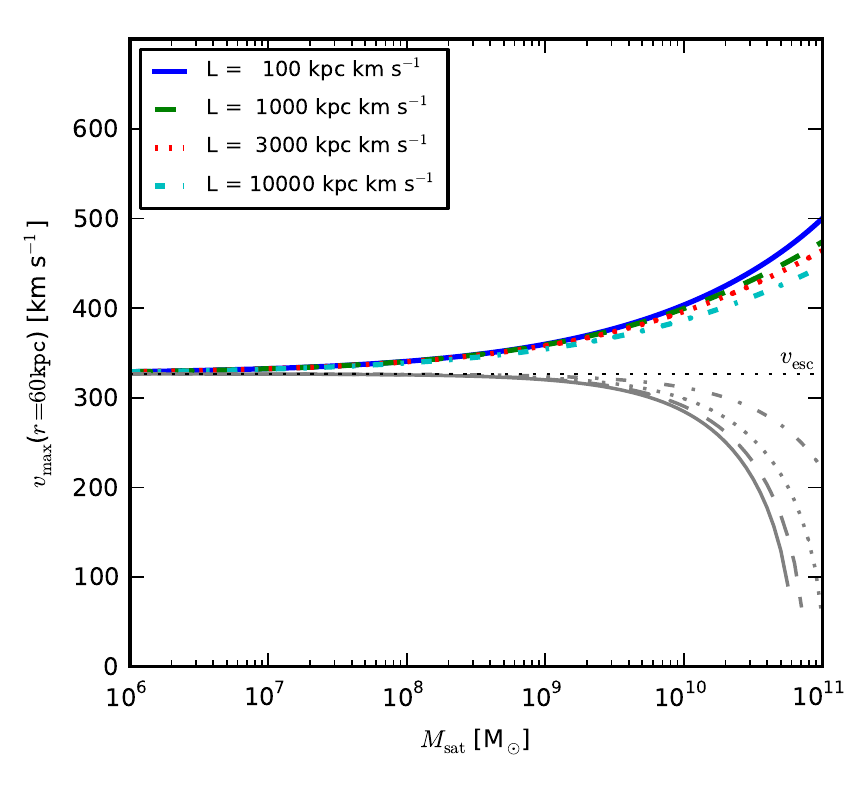}
  \caption{The maximum ejection velocities at a galactocentric distance of 60 kpc as a function of initial satellite mass as computed from Eq.~\ref{eq:vMax} assuming an initial orbital energy of the satellite $E_{\rm sat} = 0$ km$^2$s$^{-2}$. The energy loss due to dynamical friction was computed using the empirical law (Eq.~\ref{eq:dE_DF}) obtained from our simulations. The lower grey lines show the velocity of the satellite remnant at the same distance.}
  \label{fig:MaxVelocities}
\end{figure}
It is now straight forward to compute the maximum velocities generated during a tidal collision at certain galactocentric radius $r>R_{\rm peri}$:
\begin{eqnarray} \label{eq:vMax}
  v_{\rm max}(r) &=& \left[{2(E_{\rm max}-\Phi_{\rm host}(r))}\right]^\frac{1}{2}  \nonumber \\
                 &=& \left[{2(E_{\rm sat,apo} + 0.45\Delta E_{\rm max,c} - \Phi_{\rm host}(r))}\right]^\frac{1}{2},
\end{eqnarray}
where $E_{\rm sat,apo} = E_{\rm sat} - \Delta E_{\rm DF}$ is the orbital energy of the remaining satellite after the passage.
To obtain a quantitative idea we assume again a Galactocentric distance of $r=60$~kpc to be comparable to the observations of HVSs. For the energy loss via dynamical friction $\Delta E_{\rm DF}$ for simplicity we use the loss law (Eq.~\ref{eq:dE_DF}) found in our simulations. The velocities obtained in this way are plotted in Fig.~\ref{fig:MaxVelocities} as a function of the initial satellite mass $M_{\rm sat}$ and for four different initial angular momenta $L_{\rm sat}$. The satellite system was assumed to approach the host galaxy on a parabolic orbit, i.e. $E_{\rm sat} = 0$~km$^2$~s$^{-2}$. Only the most massive satellite galaxies eject HVSs with substantial velocities comparable to the radial velocities of the observed HVSs. Less massive galaxies could, in principal, also yield such large velocities if they move on more energetic orbits themselves. For example, a satellite with mass $M_{\rm sat} = 10^9$~M$_\odot$ would have to cross the virial radius of our parent galaxy\footnote{In our Galaxy model $R_{200} \simeq 260$~kpc.} with a velocity of $\sim 660$~km~s$^{-1}$ to eject a star with 720~km~s$^{-1}$ at 60~kpc comparable to the fastest HVSs known. Such a system would not lose enough energy to become bound to the larger galaxy. In this case one hardly would speak about an ejected star since the galaxy would move along with the star for a long period of time.\\
We now consider the subsample of observed HVSs with travel times $\sim 133$~Myr pointed out by \citet[cf. their Fig.~1]{Abadi2009}. The spread in velocities is roughly\footnote{We ignore the fact that the stars reside at different galactocentric radii. Taking this into account would enlarge the spread even further.} $400$ km s$^{-1}$. Such strong variations in the velocities translate into a progenitor mass of $\simeq 10^{11}$ M$_\odot$. If we select only stars within the overdensity region defined by \citet{Abadi2009} the spread reduces to $\sim 250$ km s$^{-1}$ resulting in a minimum progenitor mass still larger than $10^{10}$ M$_{\odot}$. Concerning the known satellite galaxies the respective authors of the HVS discovery papers already excluded a kinematical connection to them \citep{Brown2005, Brown2006a, Brown2009a, Edelmann2005, Hirsch2005}. Since we do not expect such a system to have escaped observations to date we conclude that a satellite origin for the subsample to be unlikely.\\
%
However, one should keep in mind that a more massive host galaxy (our host model has a total mass of~$1.1\times 10^{12}$~M$_\odot$) would shift the lines in Fig.~\ref{fig:MaxVelocities} upwards and would, in principal, allow also very small galaxies to produce stars with velocities, e.g.~$>500$~km~s$^{-1}$. However, the fact that only massive galaxies can eject stars with velocities significantly larger than their own velocities remains unaffected by this.
\paragraph{The bound HVS population and the outer stellar halo}
Several recent studies have shown that the outer stellar halo is almost purely made of accreted stars \citep{Abadi2006, Zolotov2009, Scannapieco2009}. As smooth gas accretion via cold flows plays only a minor role for Milky Way-type galaxies at low red shifts \citep{Brooks2009}, we can assume the Galaxy has not grown significantly since its last major merger.  Our simulations now demonstrate that satellite accretion will inevitably produce stars with velocities up to and exceeding their local escape speed. This means that the phase space distribution of stellar halo stars reaches all velocities up to the local escape speed at all times. For example, \citet{Smith2007} used this as a critical assumption for their technique to estimate the mass of the Galaxy.\\
However, this also means that a classification of a star as a HVS ejected from the SMBH based on its velocity is only valid for extremely large velocities. Without a confirmation of their young ages the ``bound'' HVS population in the compilation of \citet{Brown2009a} is indistinguishable from the normal (accreted) stellar halo population. To date only three stars in the survey have clear spectroscopic identification as main sequences B stars \citep{Fuentes2006, Lopez-Morales2008, Przybilla2008} while others could be old blue stragglers or blue horizontal branch stars \citep{Perets2009}.
\paragraph{An intragalactic stellar population}
\begin{figure}
  \centering
  \includegraphics{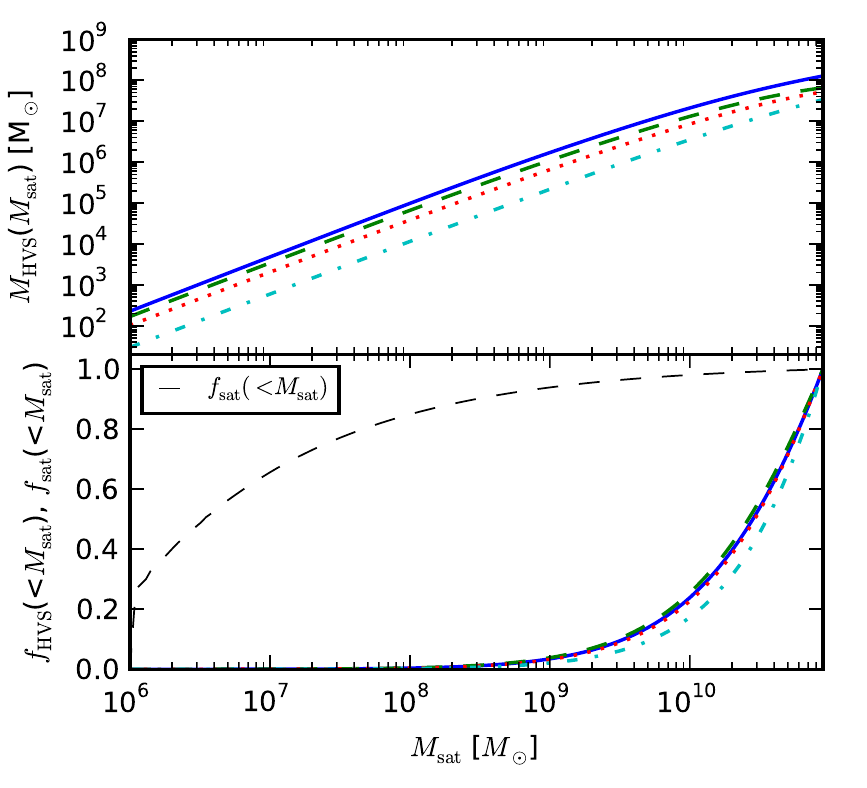}
  \caption{\textit{Upper panel}: Mass ejected at unbound velocities during one satellite orbit as a function of initial progenitor mass. The satellite is assumed to have zero initial orbital energy, i.e. to be on an parabolic orbit. Different lines correspond to four different initial angular momenta of the satellite. The vertical dashed line indicates the mass of a single star particle in our simulations. \textit{Lower panel}: mass fraction of an unbound intergalactic population originating from satellites with masses below $M_{\rm sat}$. The fraction were computed using satellite mass function based on the satellite luminosity function of \citet{Koposov2008}. The mass function is also shown as thin dashed line. More than 95 percent of the population is created by satellites more massive than $10^9$~M$_\odot$.}
  \label{fig:cumHVSmassFunc}
\end{figure}

With our results we can also address the question on what kind of satellites are the main contributors to a possible intragalactic stellar population (ISP) or Wandering stars \citep{Teyssier2009}. For this we assume the in-falling satellite galaxies to be initially on parabolic orbits, i.e. $E_{\rm sat} = 0$. Via dynamical friction the satellites will be shifted onto bound orbits during their first passage. Thus not all stars in the trailing (high energy) tidal arm will become unbound, but only those which gained more energy than is lost by their progenitor system.\\
By integrating our fitting formula (Eq.~\ref{eq:fitting_function}) using the proper value for $\epsilon_{\rm w}$ obtained via Equations~\ref{eq:correcteddEmax} and \ref{eq:dEmax-A0} over energies larger than the frictional energy loss $\Delta E_{\rm DF}$ we obtain the fraction of the baryonic mass which became HVSs. In the upper panel of Fig.~\ref{fig:cumHVSmassFunc} this fraction multiplied by the baryonic mass content of the satellite is shown as a function of the total satellite mass. For the estimation an initial angular momentum $L_0$ had to be set. The plot shows the results for four different $L_0$ (color coding is the same as in Fig.~\ref{fig:MaxVelocities}).

We then convolve this mass ejection function with an observational mass function of dwarf galaxies. We therefore used the luminosity function obtained by \citet{Koposov2008} and converted it into a mass function using the same relations used to create our satellite models for the simulations. As a result we obtained the cumulative HVS mass production function shown in the lower panel of Fig.~\ref{fig:cumHVSmassFunc}. Color coding is the same as in the upper panel. The dashed black line indicates the cumulative mass function of satellite galaxies. Consistent with the results of \citet{Teyssier2009} obtained via cosmological simulations we find that a tiny minority of massive satellites produces the overwhelming majority of HVSs. Given the fact that more massive galaxies usually are also more metal-rich \citep{Tamura2001} we conclude that an intragalactic stellar population should have at least on average a higher metalicities than the surviving dwarf galaxy population.
%
%
\section{Summary} \label{sec:Summary}
In the present work we have used a suite of 41 $N$-body simulations to study the tidal debris of satellite galaxies interacting with their much more massive host systems. \citet{Abadi2009} suggested that a fraction of the stripped-off stars can reach significant velocities and could be confound with Hypervelocity stars (HVSs) ejected from the Galactic center by a supermassive black hole. We find that, as suggested by these authors, the stripped-off stars are in fact observed in a confined region on the sky. However, for stars at distances still observable from the solar position the reason for this is not only the projection of a collimated stellar stream along the line of sight, but in addition that so shortly after the stripping event the stars had not yet enough time to disperse in physical space.

We further developed a simple analytic model to predict the maximum possible ejection velocities via estimating the maximum possible energy kick a star can obtain during such a tidal encounter (Eq.~\ref{eq:correcteddEmax}).

Following \citet{Johnston1998} we suggest that the general shape of the energy distribution of particles stripped-off during one orbit is self-similar and can be described quite accurately by Equation~\ref{eq:fitting_function}. There are only two free parameters in the distribution, its width and its normalization. The first represent a characteristic energy and is tightly connected to the maximum energy kick described by our stripping model. The normalization simply reflects the fraction of mass lost by the satellite. Both can be predicted knowing only the initial properties of the host and satellite galaxy without the need of computationally expensive $N$-body simulations.

We also address the recently reported Hypervelocity star population. Velocities larger than 500~km~s$^{-1}$ are only generated by massive satellite galaxies ($> 10^{11}$ M$_\odot$) or by galaxies with very large infall velocities in which case these galaxies stay unbound from the host and leave the parent galaxy together with the HVSs. Furthermore the larger spread in velocities of HVSs with common traveltimes also requires a massive progenitor ($> 10^{10}$ M$_\odot$). The absence of the remnant of such a massive system makes a tidal debris origin for the HVSs unlikely even from a kinematical point of view.

Convolving our formalism with a satellite mass function allows us to determine the masses of the progenitors of the main contributors to a potential intergalactic stellar population (ISP). We find that stars originating from satellite galaxies with masses $>10^9$~M$_\odot$ form about 95 percent of the population. This is consistent with the findings of \citet{Teyssier2009} who traced back the origin of unbound particles in the cosmological simulations of \citet{Bullock2005}. We conclude thus that such an ISP should tend to have the same or even a higher metallicity than the outer halo population and also as the present population of Milky Way satellites.
%
%
%
\bibliographystyle{aa}
\bibliography{/home/til/all_references}

\begin{thebibliography}{64}
\expandafter\ifx\csname natexlab\endcsname\relax\def\natexlab#1{#1}\fi

\bibitem[{{Abadi} {et~al.}(2006){Abadi}, {Navarro}, \& {Steinmetz}}]{Abadi2006}
{Abadi}, M.~G., {Navarro}, J.~F., \& {Steinmetz}, M. 2006, \mnras, 365, 747

\bibitem[{{Abadi} {et~al.}(2009){Abadi}, {Navarro}, \& {Steinmetz}}]{Abadi2009}
{Abadi}, M.~G., {Navarro}, J.~F., \& {Steinmetz}, M. 2009, \apjl, 691, L63

\bibitem[{{Boylan-Kolchin} {et~al.}(2011){Boylan-Kolchin}, {Besla}, \&
  {Hernquist}}]{Boylan-Kolchin2011}
{Boylan-Kolchin}, M., {Besla}, G., \& {Hernquist}, L. 2011, \mnras, 414, 1560

\bibitem[{{Brooks} {et~al.}(2009){Brooks}, {Governato}, {Quinn}, {Brook}, \&
  {Wadsley}}]{Brooks2009}
{Brooks}, A.~M., {Governato}, F., {Quinn}, T., {Brook}, C.~B., \& {Wadsley}, J.
  2009, \apj, 694, 396

\bibitem[{{Brown} {et~al.}(2009{\natexlab{a}}){Brown}, {Geller}, \&
  {Kenyon}}]{Brown2009a}
{Brown}, W.~R., {Geller}, M.~J., \& {Kenyon}, S.~J. 2009{\natexlab{a}}, \apj,
  690, 1639

\bibitem[{{Brown} {et~al.}(2009{\natexlab{b}}){Brown}, {Geller}, {Kenyon}, \&
  {Bromley}}]{Brown2009b}
{Brown}, W.~R., {Geller}, M.~J., {Kenyon}, S.~J., \& {Bromley}, B.~C.
  2009{\natexlab{b}}, \apjl, 690, L69

\bibitem[{{Brown} {et~al.}(2005){Brown}, {Geller}, {Kenyon}, \&
  {Kurtz}}]{Brown2005}
{Brown}, W.~R., {Geller}, M.~J., {Kenyon}, S.~J., \& {Kurtz}, M.~J. 2005,
  \apjl, 622, L33

\bibitem[{{Brown} {et~al.}(2006{\natexlab{a}}){Brown}, {Geller}, {Kenyon}, \&
  {Kurtz}}]{Brown2006a}
{Brown}, W.~R., {Geller}, M.~J., {Kenyon}, S.~J., \& {Kurtz}, M.~J.
  2006{\natexlab{a}}, \apjl, 640, L35

\bibitem[{{Brown} {et~al.}(2006{\natexlab{b}}){Brown}, {Geller}, {Kenyon}, \&
  {Kurtz}}]{Brown2006b}
{Brown}, W.~R., {Geller}, M.~J., {Kenyon}, S.~J., \& {Kurtz}, M.~J.
  2006{\natexlab{b}}, \apj, 647, 303

\bibitem[{{Brown} {et~al.}(2007{\natexlab{a}}){Brown}, {Geller}, {Kenyon},
  {Kurtz}, \& {Bromley}}]{Brown2007a}
{Brown}, W.~R., {Geller}, M.~J., {Kenyon}, S.~J., {Kurtz}, M.~J., \& {Bromley},
  B.~C. 2007{\natexlab{a}}, \apj, 660, 311

\bibitem[{{Brown} {et~al.}(2007{\natexlab{b}}){Brown}, {Geller}, {Kenyon},
  {Kurtz}, \& {Bromley}}]{Brown2007b}
{Brown}, W.~R., {Geller}, M.~J., {Kenyon}, S.~J., {Kurtz}, M.~J., \& {Bromley},
  B.~C. 2007{\natexlab{b}}, \apj, 671, 1708

\bibitem[{{Bullock} \& {Johnston}(2005)}]{Bullock2005}
{Bullock}, J.~S. \& {Johnston}, K.~V. 2005, \apj, 635, 931

\bibitem[{{Caldwell} {et~al.}(2010){Caldwell}, {Morrison}, {Kenyon},
  {Schiavon}, {Harding}, \& {Rose}}]{Caldwell2010}
{Caldwell}, N., {Morrison}, H., {Kenyon}, S.~J., {et~al.} 2010, \aj, 139, 372

\bibitem[{{Chandrasekhar}(1943)}]{Chandrasekhar1943}
{Chandrasekhar}, S. 1943, \apj, 97, 255

\bibitem[{{Choi} {et~al.}(2007){Choi}, {Weinberg}, \& {Katz}}]{Choi2007}
{Choi}, J., {Weinberg}, M.~D., \& {Katz}, N. 2007, \mnras, 381, 987

\bibitem[{{Choi} {et~al.}(2009){Choi}, {Weinberg}, \& {Katz}}]{Choi2009}
{Choi}, J., {Weinberg}, M.~D., \& {Katz}, N. 2009, \mnras, 400, 1247

\bibitem[{{de Rijcke} {et~al.}(2005){de Rijcke}, {Michielsen}, {Dejonghe},
  {Zeilinger}, \& {Hau}}]{deRijcke2005}
{de Rijcke}, S., {Michielsen}, D., {Dejonghe}, H., {Zeilinger}, W.~W., \&
  {Hau}, G.~K.~T. 2005, \aap, 438, 491

\bibitem[{{Dehnen} \& {Binney}(1998)}]{Dehnen1998}
{Dehnen}, W. \& {Binney}, J. 1998, \mnras, 294, 429

\bibitem[{{D'Onghia} {et~al.}(2009){D'Onghia}, {Besla}, {Cox}, \&
  {Hernquist}}]{DOnghia2009}
{D'Onghia}, E., {Besla}, G., {Cox}, T.~J., \& {Hernquist}, L. 2009, \nat, 460,
  605

\bibitem[{{D'Onghia} {et~al.}(2010){D'Onghia}, {Vogelsberger},
  {Faucher-Giguere}, \& {Hernquist}}]{DOnghia2010}
{D'Onghia}, E., {Vogelsberger}, M., {Faucher-Giguere}, C.-A., \& {Hernquist},
  L. 2010, \apj, 725, 353

\bibitem[{{Edelmann} {et~al.}(2005){Edelmann}, {Napiwotzki}, {Heber},
  {Christlieb}, \& {Reimers}}]{Edelmann2005}
{Edelmann}, H., {Napiwotzki}, R., {Heber}, U., {Christlieb}, N., \& {Reimers},
  D. 2005, \apjl, 634, L181

\bibitem[{{Fuentes} {et~al.}(2006){Fuentes}, {Stanek}, {Gaudi}, {McLeod},
  {Bogdanov}, {Hartman}, {Hickox}, \& {Holman}}]{Fuentes2006}
{Fuentes}, C.~I., {Stanek}, K.~Z., {Gaudi}, B.~S., {et~al.} 2006, \apjl, 636,
  L37

\bibitem[{{Gan} {et~al.}(2010){Gan}, {Kang}, {van den Bosch}, \&
  {Hou}}]{Gan2010}
{Gan}, J., {Kang}, X., {van den Bosch}, F.~C., \& {Hou}, J. 2010, \mnras, 408,
  2201

\bibitem[{{Gnedin} {et~al.}(2005){Gnedin}, {Gould}, {Miralda-Escud{\'e}}, \&
  {Zentner}}]{Gnedin2005}
{Gnedin}, O.~Y., {Gould}, A., {Miralda-Escud{\'e}}, J., \& {Zentner}, A.~R.
  2005, \apj, 634, 344

\bibitem[{{Guo} {et~al.}(2010){Guo}, {White}, {Li}, \&
  {Boylan-Kolchin}}]{Guo2010}
{Guo}, Q., {White}, S., {Li}, C., \& {Boylan-Kolchin}, M. 2010, \mnras, 404,
  1111

\bibitem[{{Helmi} \& {White}(2001)}]{Helmi2001}
{Helmi}, A. \& {White}, S.~D.~M. 2001, \mnras, 323, 529

\bibitem[{{Hernquist}(1990)}]{Hernquist1990}
{Hernquist}, L. 1990, \apj, 356, 359

\bibitem[{{Hernquist}(1993)}]{Hernquist1993}
{Hernquist}, L. 1993, \apjs, 86, 389

\bibitem[{{Hills}(1988)}]{Hills1988}
{Hills}, J.~G. 1988, \nat, 331, 687

\bibitem[{{Hirsch} {et~al.}(2005){Hirsch}, {Heber}, {O'Toole}, \&
  {Bresolin}}]{Hirsch2005}
{Hirsch}, H.~A., {Heber}, U., {O'Toole}, S.~J., \& {Bresolin}, F. 2005, \aap,
  444, L61

\bibitem[{{Johnston}(1998)}]{Johnston1998}
{Johnston}, K.~V. 1998, \apj, 495, 297

\bibitem[{{Klypin} {et~al.}(2002){Klypin}, {Zhao}, \&
  {Somerville}}]{Klypin2002}
{Klypin}, A., {Zhao}, H., \& {Somerville}, R.~S. 2002, \apj, 573, 597

\bibitem[{{Koposov} {et~al.}(2008){Koposov}, {Belokurov}, {Evans}, {Hewett},
  {Irwin}, {Gilmore}, {Zucker}, {Rix}, {Fellhauer}, {Bell}, \&
  {Glushkova}}]{Koposov2008}
{Koposov}, S., {Belokurov}, V., {Evans}, N.~W., {et~al.} 2008, \apj, 686, 279

\bibitem[{{Kregel} {et~al.}(2002){Kregel}, {van der Kruit}, \& {de
  Grijs}}]{Kregel2002}
{Kregel}, M., {van der Kruit}, P.~C., \& {de Grijs}, R. 2002, \mnras, 334, 646

\bibitem[{{Law} {et~al.}(2009){Law}, {Majewski}, \& {Johnston}}]{Law2009}
{Law}, D.~R., {Majewski}, S.~R., \& {Johnston}, K.~V. 2009, \apjl, 703, L67

\bibitem[{{Levin}(2006)}]{Levin2006}
{Levin}, Y. 2006, \apj, 653, 1203

\bibitem[{{L{\'o}pez-Morales} \& {Bonanos}(2008)}]{Lopez-Morales2008}
{L{\'o}pez-Morales}, M. \& {Bonanos}, A.~Z. 2008, \apjl, 685, L47

\bibitem[{{Lu} {et~al.}(2010){Lu}, {Zhang}, \& {Yu}}]{Lu2010}
{Lu}, Y., {Zhang}, F., \& {Yu}, Q. 2010, \apj, 709, 1356

\bibitem[{{McGaugh} {et~al.}(2010){McGaugh}, {Schombert}, {de Blok}, \&
  {Zagursky}}]{McGaugh2010}
{McGaugh}, S.~S., {Schombert}, J.~M., {de Blok}, W.~J.~G., \& {Zagursky}, M.~J.
  2010, \apjl, 708, L14

\bibitem[{{Miyamoto} \& {Nagai}(1975)}]{Miyamoto1975}
{Miyamoto}, M. \& {Nagai}, R. 1975, \pasj, 27, 533

\bibitem[{{Navarro} {et~al.}(1997){Navarro}, {Frenk}, \& {White}}]{NFW1997}
{Navarro}, J.~F., {Frenk}, C.~S., \& {White}, S.~D.~M. 1997, \apj, 490, 493

\bibitem[{{Perets} {et~al.}(2007){Perets}, {Hopman}, \&
  {Alexander}}]{Perets2007}
{Perets}, H.~B., {Hopman}, C., \& {Alexander}, T. 2007, \apj, 656, 709

\bibitem[{{Perets} {et~al.}(2009){Perets}, {Wu}, {Zhao}, {Famaey}, {Gentile},
  \& {Alexander}}]{Perets2009}
{Perets}, H.~B., {Wu}, X., {Zhao}, H.~S., {et~al.} 2009, \apj, 697, 2096

\bibitem[{{Plummer}(1911)}]{Plummer1911}
{Plummer}, H.~C. 1911, \mnras, 71, 460

\bibitem[{{Przybilla} {et~al.}(2008){Przybilla}, {Nieva}, {Tillich}, {Heber},
  {Butler}, \& {Brown}}]{Przybilla2008}
{Przybilla}, N., {Nieva}, M.~F., {Tillich}, A., {et~al.} 2008, \aap, 488, L51

\bibitem[{{Przybilla} {et~al.}(2010{\natexlab{a}}){Przybilla}, {Tillich},
  {Heber}, \& {Scholz}}]{Pryzbilla2010}
{Przybilla}, N., {Tillich}, A., {Heber}, U., \& {Scholz}, R.
  2010{\natexlab{a}}, \apj, 718, 37

\bibitem[{{Przybilla} {et~al.}(2010{\natexlab{b}}){Przybilla}, {Tillich},
  {Heber}, \& {Scholz}}]{Przybilla2010}
{Przybilla}, N., {Tillich}, A., {Heber}, U., \& {Scholz}, R.-D.
  2010{\natexlab{b}}, \apj, 718, 37

\bibitem[{{Scannapieco} {et~al.}(2011){Scannapieco}, {White}, {Springel}, \&
  {Tissera}}]{Scannapieco2011}
{Scannapieco}, C., {White}, S., {Springel}, V., \& {Tissera}, P. 2011, \mnras

\bibitem[{{Scannapieco} {et~al.}(2009){Scannapieco}, {White}, {Springel}, \&
  {Tissera}}]{Scannapieco2009}
{Scannapieco}, C., {White}, S.~D.~M., {Springel}, V., \& {Tissera}, P.~B. 2009,
  \mnras, 396, 696

\bibitem[{{Sesana} {et~al.}(2009){Sesana}, {Madau}, \& {Haardt}}]{Sesana2009}
{Sesana}, A., {Madau}, P., \& {Haardt}, F. 2009, \mnras, 392, L31

\bibitem[{{Smith} {et~al.}(2007){Smith}, {Ruchti}, {Helmi}, {Wyse},
  {Fulbright}, {Freeman}, {Navarro}, {Seabroke}, {Steinmetz}, {Williams},
  {Bienaym{\'e}}, {Binney}, {Bland-Hawthorn}, {Dehnen}, {Gibson}, {Gilmore},
  {Grebel}, {Munari}, {Parker}, {Scholz}, {Siebert}, {Watson}, \&
  {Zwitter}}]{Smith2007}
{Smith}, M.~C., {Ruchti}, G.~R., {Helmi}, A., {et~al.} 2007, \mnras, 379, 755

\bibitem[{{Springel}(2005)}]{GadgetPaper}
{Springel}, V. 2005, \mnras, 364, 1105

\bibitem[{{Springel} \& {White}(1999)}]{SW1999}
{Springel}, V. \& {White}, S.~D.~M. 1999, \mnras, 307, 162

\bibitem[{{Tamura} {et~al.}(2001){Tamura}, {Hirashita}, \&
  {Takeuchi}}]{Tamura2001}
{Tamura}, N., {Hirashita}, H., \& {Takeuchi}, T.~T. 2001, \apjl, 552, L113

\bibitem[{{Taylor} \& {Babul}(2001)}]{Taylor2001}
{Taylor}, J.~E. \& {Babul}, A. 2001, \apj, 559, 716

\bibitem[{{Teuben}(1995)}]{NEMOPaper}
{Teuben}, P. 1995, in Astronomical Society of the Pacific Conference Series,
  Vol.~77, Astronomical Data Analysis Software and Systems IV, ed. {R.~A.~Shaw,
  H.~E.~Payne, \& J.~J.~E.~Hayes}, 398--+

\bibitem[{{Teyssier} {et~al.}(2009){Teyssier}, {Johnston}, \&
  {Shara}}]{Teyssier2009}
{Teyssier}, M., {Johnston}, K.~V., \& {Shara}, M.~M. 2009, \apjl, 707, L22

\bibitem[{{Tillich} {et~al.}(2009){Tillich}, {Przybilla}, {Scholz}, \&
  {Heber}}]{Tillich2009}
{Tillich}, A., {Przybilla}, N., {Scholz}, R., \& {Heber}, U. 2009, \aap, 507,
  L37

\bibitem[{{Tormen} {et~al.}(1998){Tormen}, {Diaferio}, \& {Syer}}]{Tormen1998}
{Tormen}, G., {Diaferio}, A., \& {Syer}, D. 1998, \mnras, 299, 728

\bibitem[{{Warnick} {et~al.}(2008){Warnick}, {Knebe}, \& {Power}}]{Warnick2008}
{Warnick}, K., {Knebe}, A., \& {Power}, C. 2008, \mnras, 385, 1859

\bibitem[{{Xue} {et~al.}(2008){Xue}, {Rix}, {Zhao}, {Re Fiorentin}, {Naab},
  {Steinmetz}, {van den Bosch}, {Beers}, {Lee}, {Bell}, {Rockosi}, {Yanny},
  {Newberg}, {Wilhelm}, {Kang}, {Smith}, \& {Schneider}}]{Xue2008}
{Xue}, X.~X., {Rix}, H.~W., {Zhao}, G., {et~al.} 2008, \apj, 684, 1143

\bibitem[{{Yu} \& {Madau}(2007)}]{Yu2007}
{Yu}, Q. \& {Madau}, P. 2007, \mnras, 379, 1293

\bibitem[{{Yu} \& {Tremaine}(2003)}]{Yu2003}
{Yu}, Q. \& {Tremaine}, S. 2003, \apj, 599, 1129

\bibitem[{{Zolotov} {et~al.}(2009){Zolotov}, {Willman}, {Brooks}, {Governato},
  {Brook}, {Hogg}, {Quinn}, \& {Stinson}}]{Zolotov2009}
{Zolotov}, A., {Willman}, B., {Brooks}, A.~M., {et~al.} 2009, \apj, 702, 1058

\end{thebibliography}
%
%
%
%
\appendix
\section{Estimating the maximum energy gain} \label{app:EnergyEstimate}
In this section we briefly describe the process of how we obtained estimates for the velocities $V_{\rm peri}$ and $v_{\rm esc}(R_{\rm tidal})$ which we need to evaluate Equation~\ref{eq:deltaE}. The ingredients for this are
\begin{itemize}
  \item the radial mass profile of the host galaxy,
  \item the radial dark matter and baryonic mass profile of the satellite galaxy,
  \item the parameters of the satellite orbit, namely the initial angular momentum $L_{\rm sat}$ and the initial orbital energy $E_{\rm sat}$.
\end{itemize}
In a first step we estimate the minimum distance to which the satellite approaches the host center, i.e. the pericenter distance $R_{\rm peri}$. For this we use the effective potential $\Phi_{\rm eff} = L_{\rm sat}^2 / (2r^2) + \Phi_{\rm host}(r)$ exploiting
\begin{equation} \label{App:eq:rperi}
  E_{\rm sat} - \frac{1}{2}\Delta E_{\rm DF} = \Phi_{\rm eff}(R_{\rm peri}),
\end{equation}
where we compute the energy loss from dynamical friction $\Delta E_{\rm DF}$ using Eq.~\ref{eq:dE_DF}. Further we compute the satellite velocity in the Perigalacticon via
\begin{equation}
  V_{\rm peri} = \sqrt{2(E_{\rm sat} - \frac{1}{2}\Delta E_{\rm DF} - \Phi_{\rm host}(R_{\rm peri}))}.
\end{equation}

To compute the escape velocity $v_{\rm esc}(R_{\rm tidal})$ from the satellite system we first have to determine the tidal radius $R_{\rm tidal}$ which we assume to be equal to the Jacobi radius at the distance $R_{\rm peri}$:
\begin{equation} \label{eq:Rtidal}
 R_{\rm tidal} = \left(\frac{M'_{\rm sat}}{3M_{\rm host}(R_{\rm peri})} \right)^{\frac{1}{3}} R_{\rm peri}.
\end{equation}
However, we do not take the total satellite mass $M_{\rm sat}$ for the final radius. We also take into account that due to its much larger extension the dark matter halo of the satellite is stripped much earlier the the baryonic component. Consequently in a first step we compute the tidal radius using the total satellite mass $M_{\rm sat}$ and assume that all material outside this ``dark matter tidal radius'' $R_{\rm tidal, DM}$ is lost. We then compute the ``baryonic tidal radius'' using Equation~\ref{eq:Rtidal} with the mass $M'_{\rm sat} = M_{\rm sat}(r<R_{\rm tidal, DM})$.\\
Finally we obtain the escape speed
\begin{equation}
  v_{\rm esc}(R_{\rm tidal}) = \sqrt{2\vert \Phi_{\rm sat}(R_{\rm tidal})\vert}
\end{equation}
Note, that the tidal radius computed in this two-step process also allows a very good estimate of the baryonic mass loss of the satellite when it is assumed that all mass outside this tidal radius is lost, i.e.
\begin{equation}
 f_{\rm unbound} = \dfrac{M_{\rm lost}}{M_{\rm sat}} = \dfrac{M_{\rm sat}(r>R_{\rm tidal})}{M_{\rm sat}}
\end{equation}
This was used in Sect.~\ref{sec:Discussion} to estimate the fraction of satellite mass expelled as HVSs into intergalactic space.
%
%
%
%
\section{Scaling tests} \label{app:Scaling-Tests}
\begin{figure}
  \centering
  \includegraphics{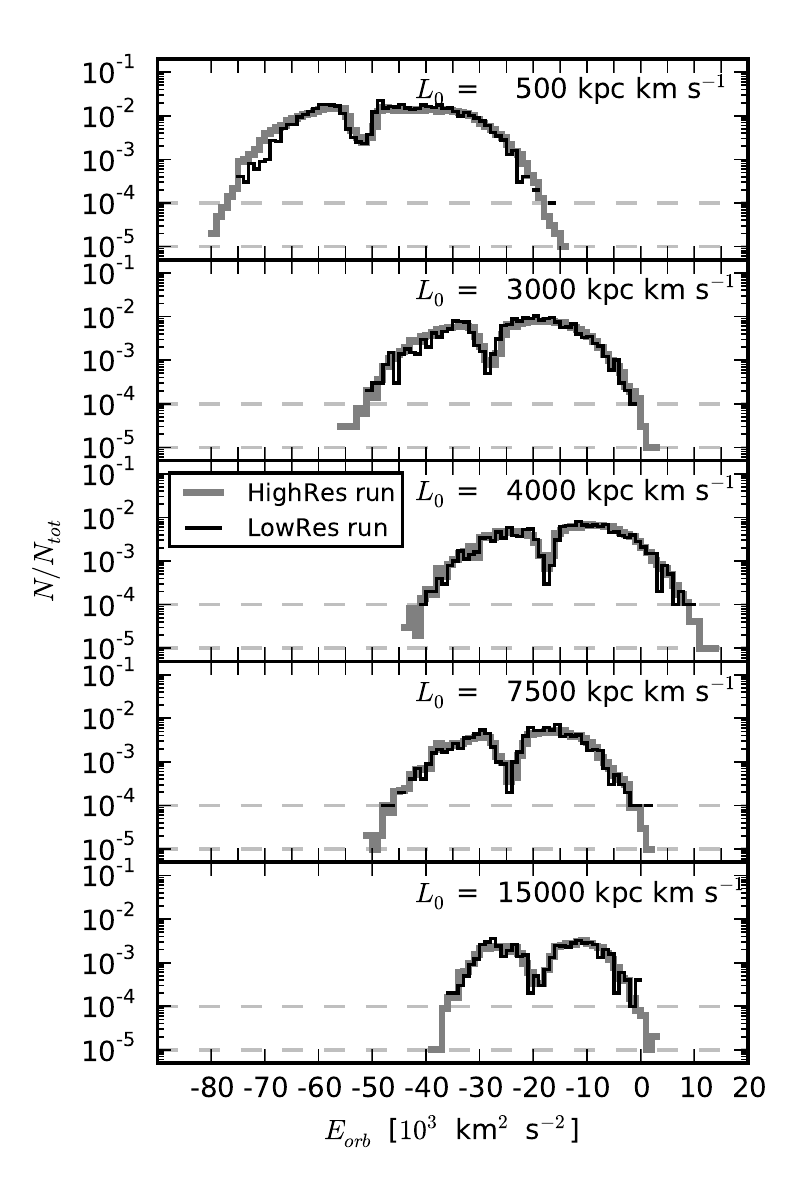}
  \caption{Comparison of the energy distributions obtained from corresponding high and low resolution runs. The dashed lines indicate the mass resolution limits of the simulations, i.e. the mass of a single star particle.}
  \label{fig:ScTdist}
\end{figure}
To assess the influence of the numerical resolution on our results we ran a set of simulations with the number of particles in the satellite system reduced by a factor of 10. This was done by taking the initial snapshot file of one of our high resolution runs and randomly removing 90 percent of the satellite particles. Then the masses of the remaining particles increased by a factor of 10. In this way we obtain an equilibrium configuration of a satellite system with exactly the same properties as the high resolution one.\\
In Figure~\ref{fig:ScTdist} we show the resulting energy distributions together with those of the corresponding high resolution runs with the same initial conditions. Dashed lines indicate the respective mass resolution limits, i.e. the mass of a single star particle in the simulations. The shapes of the distributions show no significant differences. The width of the corresponding distributions, $\epsilon_{\rm w}$, obtained by fitting Eq.~\ref{eq:fitting_function} also differ by less than 10 percent. As could be expected the low resolution distribution does not reach as high energies as the high resolution one. However, the maximum energies do not differ by much, due to the steep slope in the outer tails of the distribution.

We also repeated one simulation run using a 5 times larger softening length for the star particles. For all quantities measured for this study the outcome changed by less than 1 percent. Especially the maximum energies reached by satellite particles differ only by 0.1 percent. This shows that our results are not affected by artificial heating by two-body encounters.
%
%
\section{Other host galaxy potentials} \label{App:sec:hostPot_tests}
\begin{figure}
 \centering
 \includegraphics{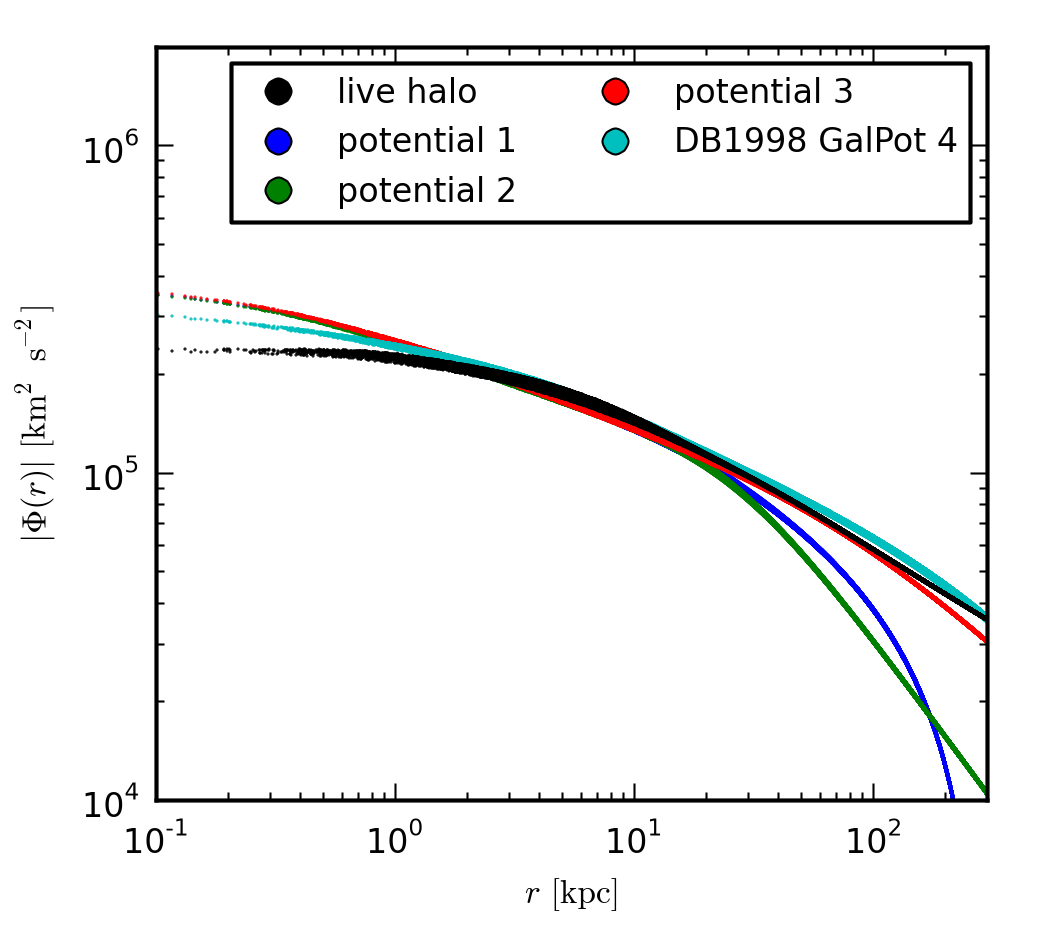}
 \caption{Radial gravitational potential profiles of the four alternative host galaxy representation and for the $N$-body live host galaxy (black). The alternative models cover a variety of central and outer slopes allowing to test the influence of those on satellite tidal tails. The varying thickness of the profile lines reflects the non-spherical symmetry of the potentials.}
 \label{App:fig:comparisonHostPot}
\end{figure}
\begin{figure}
 \centering
 \includegraphics{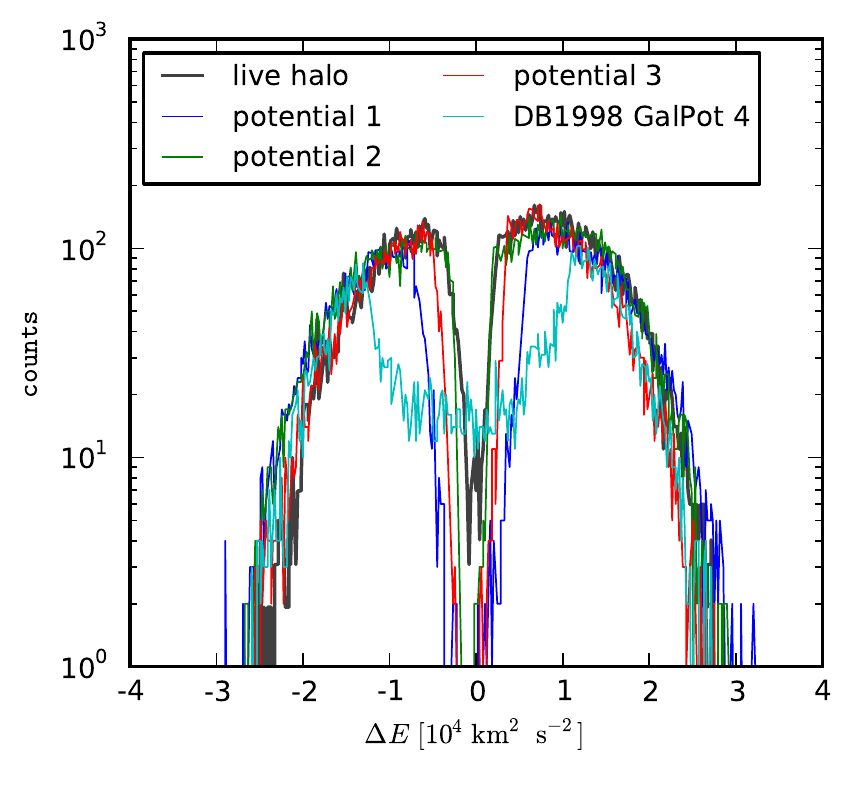}
 \caption{Comparison of the energy distributions of the tidal tail particle stripped from identical satellite galaxies with identical initial phase space positions evolving in different host potentials. The outer tails are virtually unaffected by the differing potentials, while the central minimum is subject to significant changes.}
 \label{App:fig:comparisonEdist}
\end{figure}
To test the influence of the shape of the Milky Way-host potential we performed a small number of test simulations with different rigid potentials representing the host galaxy. Four different models were applied: three of them (potentials 1--3) share the same parametrization of the baryonic disk and spheriod components. A \citet{Miyamoto1975} disk with mass $10^{11}~{\rm M_{\odot}}$, radial and vertical scale length of 6.5 and 0.26 kpc, respectively and a \citet{Hernquist1990} bulge component of mass $3.4\times 10^{10}~{\rm M}_\odot$ and a scale length of 0.7 kpc. The dark haloes are modeled with
\begin{itemize}
 \item \textit{potential 1}: a logarithmic potential $\Phi(r) = v_{\rm h}^2\ln(r^2+r_{\rm h}^2)$ with $v_{\rm h} = 128$ km/s and $r_{\rm h} = 12$ kpc. This potential was already often used by other studies to mimick a Milky Way \citep[e.g.][]{Johnston1998, Helmi2001}.
 \item \textit{potential 2}: a \citet{Plummer1911} sphere with mass $6\times10^{11}~{\rm M}_\odot$ and scale radius 25.7 kpc.
 \item \textit{potential 3}: a NFW sphere \citep{NFW1997} with central density $1.523\times10^6~{\rm M}_\odot{\rm kpc}^{-3}$ and scale radius 36 kpc.
\end{itemize}
All three potentials were chosen such that $V_{\rm circ}(8.5~\rm kpc) = 220$ km/s and $V_{\rm esc}(8.5~\rm kpc) = 550$ km/s. As a fourth option we used the model potential number 4 of \citet{Dehnen1998} which we implemented into Gadget-2 using a C++ routine prepared by Walter Dehnen and distributed with the NEMO Stellar Dynamics Toolbox \citep{NEMOPaper}.

Figure~\ref{App:fig:comparisonHostPot} shows a comparison of the radial profiles of the four potentials with the radial profile of the live halo used in the main part of the simulations. All potentials have a steeper slope in the inner regions. The virtually flat part of the live potential is due to the gravitational softening becoming significant on these scales.\\
Figure~\ref{App:fig:comparisonEdist} plots the energy tidal tail distributions (cf. Fig.~\ref{fig:Edist}) obtained with the different host representations but otherwise identical initial conditions. While the distribution changes strongly in regions with small $|\Delta E|$, the tail of the distribution remain virtually unchanged. We thus conclude that the actual shape of the Galactic potential has no major influence on our results. The variations around the central minimum are most likely due to different evolution of the Roche radius of the satellite during its orbits determining whether particles with low $|\Delta E|$ which stay near the satellite for longer periods are re-captured.
%
%
\section{Initial conditions}
%
%
\begin{table*}
  \centering
  \caption{Initial parameters of the satellite systems (plus some analysis results)}
  \begin{tabular}{r c c c c c c c c c}
\hline\\ [-7pt]
  No. &$N_*^1$&     $N_{DM}^2$    &   $M_{200}^3$      &     $M_{bary}^4$   &    $L_0^5$              &      $E_0^6$           & orbit type$^7$     & $\epsilon_{\rm w}^8$& $f_{ub}^9$ \\[2pt]
      &          &                 &$10^{10}$ M$_\odot$& $10^8$ M$_\odot$&$10^3$ kpc km s$^{-1}$ &$10^4$ km$^2$ s$^{-2}$&                 &$10^4$ km$^2$ s$^{-2}$ &\\
\hline\hline\\ [-8pt]
\multicolumn{10}{c}{\tt Runs with reduced resolution} \\
\hline\\ [-8pt]
  001 &  $10^4$ &  $5 \times 10^3$ & 1.0              & 1.0              &  0.5                  &  -3.8  & polar                &  2.2     &    0.49 \\    
  002 &  $10^4$ &  $5 \times 10^3$ & 1.0              & 1.0              &  3                    &  -1.8  & polar                &  1.9     &    0.30 \\    
  003 &  $10^4$ &  $5 \times 10^3$ & 1.0              & 1.0              &  4                    &  -0.9  & polar                &  1.8     &    0.15 \\    
  004 &  $10^4$ &  $5 \times 10^3$ & 1.0              & 1.0              &  7.5                  &  -1.8  & polar                &  1.6     &    0.10 \\    
  005 &  $10^4$ &  $5 \times 10^3$ & 1.0              & 1.0              &  15                   &  -1.5  & polar                &  1.7     &    0.05 \\    
\hline\\ [-8pt]
\multicolumn{10}{c}{\tt Runs with full resolution} \\
\hline\\ [-8pt]
  006 &  $10^5$ &  $5 \times 10^4$ & 0.7              & 0.8              &  0                    &  -1.8  & polar                &  2.1     &    0.65 \\    
  007 &  $10^5$ &  $5 \times 10^4$ & 1.0              & 1.0              &  0                    &  -1.8  & polar                &  2.6     &    0.56 \\    

  008 &  $10^5$ &  $5 \times 10^4$ & 0.1              & 0.06             &  0.5                  &  -3.8  & polar                &  0.6     &    0.71 \\    
  009 &  $10^5$ &  $5 \times 10^4$ & 0.3              & 0.2              &  0.5                  &  -3.8  & polar                &  1.1     &    0.58 \\    
  010 &  $10^5$ &  $5 \times 10^4$ & 0.5              & 0.5              &  0.5                  &  -3.8  & polar                &  1.6     &    0.55 \\    
  011 &  $10^5$ &  $5 \times 10^4$ & 0.5              & 0.5              &  0.5                  &  -3.8  & polar                &  1.6     &    0.55 \\    
  012 &  $10^5$ &  $5 \times 10^4$ & 0.7              & 0.8              &  0.5                  &  -3.8  & polar                &  1.9     &    0.51 \\    
  013 &  $10^5$ &  $5 \times 10^4$ & 1.0              & 1.0              &  0.5                  &  -3.8  & polar                &  2.4     &    0.47 \\    
  014 &  $10^5$ &  $5 \times 10^4$ & 2.0              & 3.0              &  0.5                  &  -3.8  & polar                &  3.7     &    0.42 \\    

  015 &  $10^5$ &  $5 \times 10^4$ & 0.1              & 0.06             &  3                    &  -1.8  & polar                &  0.7     &    0.26 \\    
  016 &  $10^5$ &  $5 \times 10^4$ & 0.3              & 0.2              &  3                    &  -1.8  & polar                &  1.1     &    0.23 \\    
  017 &  $10^5$ &  $5 \times 10^4$ & 0.5              & 0.5              &  3                    &  -1.8  & polar                &  1.5     &    0.21 \\    
  018 &  $10^5$ &  $5 \times 10^4$ & 0.7              & 0.8              &  3                    &  -1.8  & polar                &  1.7     &    0.20 \\    
  019 &  $10^5$ &  $5 \times 10^4$ & 1.0              & 1.0              &  3                    &  -1.8  & polar                &  2.0     &    0.19 \\    
  020 &  $10^5$ &  $5 \times 10^4$ & 1.0              & 1.0              &  3                    &  -1.8  & polar                &  2.0     &    0.20 \\    
  021 &  $10^5$ &  $5 \times 10^4$ & 2.0              & 3.0              &  3                    &  -1.8  & polar                &  2.9     &    0.20 \\    
  022 &  $10^5$ &  $5 \times 10^4$ & 1.0              & 1.0              &  3                    &  -1.8  & polar                &  2.0     &    0.19 \\    

  023 &  $10^5$ &  $5 \times 10^4$ & 0.3              & 0.2              &  15                   &  -1.5  & polar                &  1.0     &    0.06 \\    
  024 &  $10^5$ &  $5 \times 10^4$ & 0.5              & 0.5              &  15                   &  -1.5  & polar                &  1.2     &    0.06 \\    
  025 &  $10^5$ &  $5 \times 10^4$ & 0.7              & 0.8              &  15                   &  -1.5  & polar                &  1.4     &    0.06 \\    
  026 &  $10^5$ &  $5 \times 10^4$ & 1.0              & 1.0              &  15                   &  -1.5  & polar                &  1.6     &    0.05 \\    

  027 &  $10^5$ &  $5 \times 10^4$ & 0.5              & 0.5              &  3                    &  -2.8  & polar                &  1.4     &    0.23 \\    
  028 &  $10^5$ &  $5 \times 10^4$ & 0.5              & 0.5              &  3                    &  -1.2  & polar                &  1.4     &    0.20 \\    
  029 &  $10^5$ &  $5 \times 10^4$ & 0.5              & 0.5              &  3                    &  -0.8  & polar                &  1.4     &    0.20 \\    

  030 &  $10^5$ &  $5 \times 10^4$ & 1.0              & 1.0              &  7.5                  &  -1.7  & polar                &  1.8     &    0.10 \\    
  031 &  $10^5$ &  $5 \times 10^4$ & 1.0              & 1.0              &  7.5                  &  -1.7  & $45^\circ$ retrograde&  1.8     &    0.10 \\    
  032 &  $10^5$ &  $5 \times 10^4$ & 1.0              & 1.0              &  7.5                  &  -1.7  & $45^\circ$ prograde  &  1.8     &    0.11 \\    
  033 &  $10^5$ &  $5 \times 10^4$ & 1.0              & 1.0              &  3                    &  -1.8  & planar retrograde    &  2.1     &    0.25 \\    
  034 &  $10^5$ &  $5 \times 10^4$ & 1.0              & 1.0              &  3                    &  -1.8  & planar prograde      &  2.0     &    0.25 \\    
  035 &  $10^5$ &  $5 \times 10^4$ & 1.0              & 1.0              &  7.5                  &  -1.7  & planar retrograde    &  1.8     &    0.11 \\    
  036 &  $10^5$ &  $5 \times 10^4$ & 1.0              & 1.0              &  7.5                  &  -1.7  & planar prograde      &  1.8     &    0.12 \\    
  037 &  $10^5$ &  $5 \times 10^4$ & 0.5              & 0.5              &  15                   &  -1.5  & planar retrograde    &  1.2     &    0.06 \\    
  038 &  $10^5$ &  $5 \times 10^4$ & 0.5              & 0.5              &  15                   &  -1.5  & planar prograde      &  1.2     &    0.06 \\    

  039 &  $10^5$ &  $5 \times 10^4$ & 1.0              & 1.0              &  1                    &  -2.8  & polar                &  2.6     &    0.33 \\    
  040 &  $10^5$ &  $5 \times 10^4$ & 1.0              & 1.0              &  1.6                  &  -2.4  & polar                &  2.3     &    0.27 \\    
  041 &  $10^5$ &  $5 \times 10^4$ & 1.0              & 1.0              &  2                    &  -2.4  & polar                &  2.2     &    0.25 \\    
  042 &  $10^5$ &  $5 \times 10^4$ & 1.0              & 1.0              &  4                    &  -0.9  & polar                &  1.9     &    0.15 \\    
  043 &  $10^5$ &  $5 \times 10^4$ & 0.5              & 0.5              &  7.5                  &  -1.8  & planar retrograde    &  1.4     &    0.13 \\    

  044 &  $10^5$ &  $5 \times 10^4$ & 1.0              & 1.0              &  1                    &   0.0  & $82^\circ$ prograde  &  2.5     &    0.42 \\    
  045 &  $10^5$ &  $5 \times 10^4$ & 1.0              & 1.0              &  5                    &   0.0  & $87^\circ$ retrograde&  2.0     &    0.12 \\    
  046 &  $10^5$ &  $5 \times 10^4$ & 1.0              & 1.0              &  10                   &   0.0  & $84^\circ$ retrograde&  1.9     &    0.06 \\ [2pt] 
  \hline
  \end{tabular}
  \label{tab:ICs}
  \footnotesize
  \\ [0.5ex]
  $^1$number of star particles in the satellite\\
  $^2$number of particles in the satellite dark halo\\
  $^3$virial mass of the satellite\\
  $^4$mass of the satellite baryonic component\\
  $^5$initial angular momentum of the satellite with respect to the host galaxy\\
  $^6$initial orbital energy of the satellite\\
  $^7$inclination of the satellite orbital plane with respect to the host disk plane and,\\
  if appropriate, sense of rotation relative to the host disk rotation\\
  $^8$width of the resulting energy distribution as defined in Eq.~\ref{eq:fitting_function}\\
  $^9$fraction of baryonic mass lost during the first orbit of the satellite.
\end{table*}

\end{document}